\documentclass[draft]{agujournal2019}
\usepackage{url} 
\usepackage{amssymb,stmaryrd}
\usepackage{amsmath}
\usepackage{bm}
\usepackage{graphicx}
\usepackage{multirow}
\usepackage{xcolor}
\usepackage{textcomp}
\usepackage{float}

\usepackage[inline]{trackchanges} 
\usepackage{soul}

\draftfalse

\journalname{Journal of Advances in Modeling Earth Systems (JAMES)}

\begin{document}
\raggedbottom 

%
%

\title{Navigating the Noise: Bringing Clarity to ML Parameterization Design with $\mathcal{O}$(100) Ensembles}

%
%




\authors{Jerry Lin$^1$, Sungduk Yu$^2$, Liran Peng$^1$, Tom Beucler$^{3,4}$, Eliot Wong-Toi$^5$, Zeyuan Hu$^{6,7}$, Pierre Gentine $^{8}$, Margarita Geleta $^{9, 10}$, Mike Pritchard$^{1,7}$}

\affiliation{1}{Department of Earth System Sciences, University of California at Irvine, Irvine, CA, USA}
\affiliation{2}{Multimodal Cognitive AI, Intel Labs, Santa Clara, CA 95054, USA}
\affiliation{3}{Faculty of Geosciences and Environment, University of Lausanne, Lausanne, Switzerland}
\affiliation{4}{Expertise Center for Climate Extremes, University of Lausanne, Lausanne, Switzerland}
\affiliation{5}{Department of Statistics, University of California at Irvine, Irvine, CA, USA}
\affiliation{6}{Department of Earth and Planetary Sciences, Harvard University}
\affiliation{7}{NVIDIA Research}
\affiliation{8}{LEAP Science and Technology Center, School of Engineering and Applied Sciences, Climate School, Columbia University}
\affiliation{9}{Berkeley AI Research (BAIR), Department of Electrical Engineering and Computer Science, University of California at Berkeley, Berkeley, CA, USA}
\affiliation{10}{Department of Biomedical Data Science, Stanford University School of Medicine, Palo Alto, CA, USA}





\correspondingauthor{Jerry Lin}{jerryL9@uci.edu}



\begin{keypoints}
\item Reducing offline error can have differing effects on online error and online stability, which do not necessarily improve together.
\item Using memory, batch normalization, and training on multiple climates improve both online error and stability.
\item Ensemble sizes of $\mathcal{O}(100)$ may be necessary to detect causally relevant differences in the online performance of ML parameterizations.
\end{keypoints}

%
%

%
%


\begin{abstract}

Machine-learning (ML) parameterizations of subgrid processes (here of turbulence, convection, and radiation) may one day replace conventional parameterizations by emulating high-resolution physics without the cost of explicit simulation. However, uncertainty about the relationship between offline and online performance (i.e., when integrated with a large-scale general circulation model (GCM)) hinders their development. Much of this uncertainty stems from limited sampling of the noisy, emergent effects of upstream ML design decisions on downstream online hybrid simulation. Our work rectifies the sampling issue via the construction of a semi-automated, end-to-end pipeline for $\mathcal{O}(100)$ size ensembles of hybrid simulations, revealing important nuances in how systematic reductions in offline error manifest in changes to online error and online stability. For example, removing dropout and switching from a Mean Squared Error (MSE) to a Mean Absolute Error (MAE) loss both reduce offline error, but they have opposite effects on online error and online stability. Other design decisions, like incorporating memory, converting moisture input from specific humidity to relative humidity, using batch normalization, and training on multiple climates do not come with any such compromises. Finally, we show that ensemble sizes of $\mathcal{O}(100)$ may be necessary to reliably detect causally relevant differences online. By enabling rapid online experimentation at scale, we can empirically settle debates regarding subgrid ML parameterization design that would have otherwise remained unresolved in the noise.

\end{abstract}

\section*{Plain Language Summary}
Running high-resolution simulations of convection for climate forecasts takes a lot of computing power, which is why we use simpler and less accurate approximations for processes inside each grid cell of a climate model. Neural networks could mimic more accurate ``smaller-scale" subgrid simulations at a fraction of the computational cost, but they can also behave unpredictably when they are actually used inside the ``larger-scale" climate model (i.e., in an ``online" setting). To account for this unpredictability and understand how to make improvements with statistical confidence, we conduct thousands of ``online" experiments. We would like to reduce the online error of these models and improve their stability (i.e., preventing situations in which they crash and stop running). We found that using data from different climates and a previous time step helps with both goals, converting from specific humidity to relative humidity improves stability, and removing dropout reduces error at the cost of stability. With all this testing, we’re helping figure out how to develop neural networks that can work more reliably inside climate models, moving us closer to using them in actual climate projections.

%
%

%


%
%
%
%

\section{Introduction}
%

Despite the fact that they occur at scales smaller than the grid cell resolution of a climate model, subgrid processes (e.g. turbulence, deep convection, microphysics) can have major effects on model accuracy and fidelity. Unfortunately, resolving small-scale processes explicitly is too computationally costly for long-term climate projections. This necessitates the use of sub-grid parameterizations, i.e., empirical representation of the impact of subgrid processes on the coarse grid, which are inevitably error-prone due to their phenomenological nature. For over two decades, neural networks (NNs) have held the promise of circumventing the large amounts of compute required to resolve fine scale processes in climate simulations by learning the coarse representation of subgrid processes from high-resolution data \cite{Chevallier1998-ne, Krasnopolsky2013-ed, Gentine2018-ux, Rasp2018-fk, Yu2023-on}. Of particular interest is the potential to replace parameterizations that crudely approximate highly nonlinear subgrid scale processes like ocean momentum transport \cite{Guillaumin2021-xg}, atmospheric momentum transport via small-scale gravity waves \cite{Mansfield2024-nj}, radiative transfer \cite{Cachay2021-ds}, and moist convection \cite{Gentine2018-ux, Rasp2018-fk, Brenowitz2020-bv, Yuval2021-bh, Wang2022-po, iglesias2024causally} to create hybrid physics-ML (i.e., physical model with embedded ML parameterizations) climate models. If NNs could be trained to reliably emulate the behavior of more explicit simulation of subgrid processes, climate models could be improved compared to traditional parameterizations without incurring the associated computational cost of high-resolution simulations. In the context of convection, this could mean breaking the parameterization ``deadlock'' as coined by \citeA{Randall2003-db}---a situation in which slowing progress on traditional convective parameterizations has not kept pace with the growing societal need to accurately represent the underlying subgrid physics in climate models. \cite{Randall2013-hl, Shepherd2014-cm, Gentine2018-ux, IPCC2021}.

The vision of breaking deadlock with NN parameterizations has been challenged by difficulties ensuring their reliability when they are dynamically coupled to a host climate model (i.e., used online). NNs are typically first fit offline on some training data (typically high-resolution simulations) and then plugged online, where emergent failure modes are revealed. NNs have been shown to outperform conventional convective parameterizations when evaluated offline i.e., on test data from the cloud-resolving models (CRMs) they are trained to emulate \cite{Gentine2018-ux, Han2020-xy, Mooers2021-sh}. When coupled online, however, small offline imperfections can compound across timesteps, resulting in dramatic error growth and simulation crashes (i.e. cases where the coupled simulations prematurely terminate because of numerical instabilities) \cite{Brenowitz2020-bv, Wang2022-po}. Previous work has argued that skillful offline fits---which can be found via ample ML tools that exist to optimize it---are no guarantee for stable online performance \cite{Ott2020-qe, Wang2022-po}. 

An overarching problem is that the coupled dynamics between the GCM and NN parameterization are invisible to the optimizer during offline training, and the downstream online effects are rarely sampled at the scale required to distinguish emergent signal from noise. We hypothesize that the main obstacle is the associated technical work: training thousands of ML parameterizations, adapting each to run within a climate model, and running multi-hundred member ensembles of hybrid physics-ML climate simulations can be technically challenging and computationally intensive. As a result, basic questions such as ``what differences in online error can be statistically shown to originate from offline ML architecture and optimization choices?'' are not empirically addressed with statistical confidence. 

To answer these questions, we designed a modular, semi-automated, and end-to-end pipeline for rapidly sampling the coupled behavior of fully-connected NN emulators of moist convection using multi-hundred member ensembles. We use this capability to conduct an ablation study of different design decisions for full-physics emulators of subgrid atmospheric processes (see Section 2.3 for details).

\begin{figure}[!htbp]
 \centering
 \includegraphics[width=\textwidth]{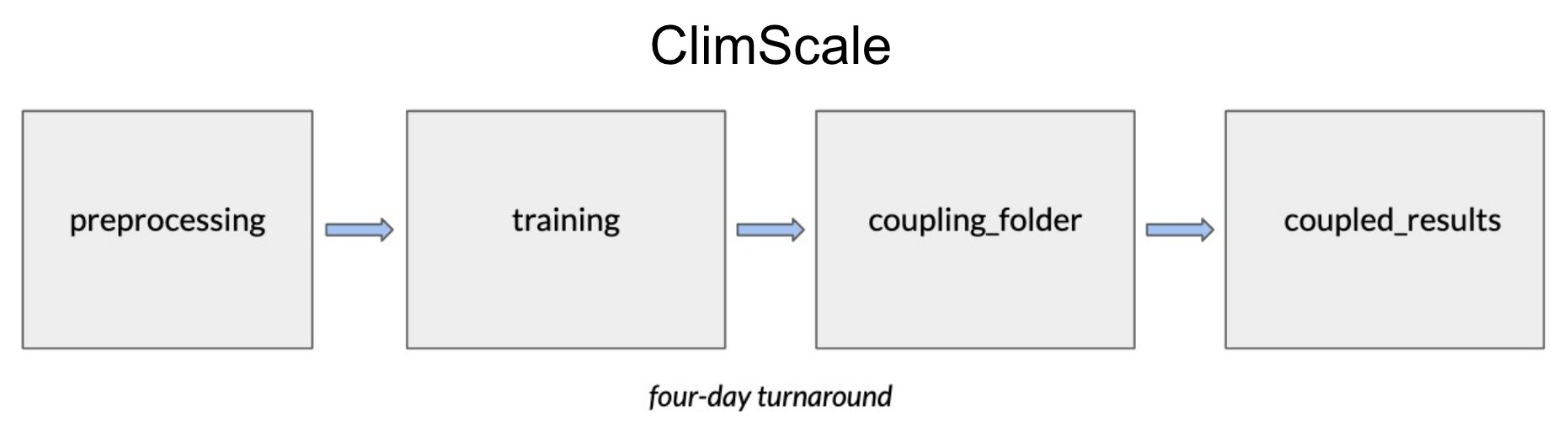}
 \setlength{\belowcaptionskip}{-1em}%

\caption{This is a diagram showcasing ClimScale, our end-to-end pipeline for going from preprocessing to online results. Associated code can be found in our GitHub repository: https://github.com/SciPritchardLab/ClimScale.}
 \label{fig:pipeline_diagram}
\end{figure}

Our level of online sampling is extensive and at least two orders of magnitude higher than what is traditionally shown in the literature \cite{Wang2022-po, Han2023-nm, iglesias2024causally}. We will show in the following sections that sampling online at a scale of hundreds of architecture trials leads to reproducible improvements to hybrid model performance that become statistically identifiable.

\section{Methods}

\subsection{Reference Climate Simulation}

The data used to train the NNs comes from the Super-Parameterized Community Atmosphere Model v3 (SPCAM 3) in an aquaplanet setting of intermediate complexity that has proved popular for exploring trade-offs in hybrid-ML simulation \cite{Gentine2018-ux, Rasp2018-fk, Ott2020-qe, Behrens2022-qq, Beucler2024-vb}. In superparameterization, each grid cell of a coarse-resolution general circulation model (GCM) contains an idealized two-dimensional CRM that explicitly resolves convection while making use of parameterizations for small-scale turbulence and cloud microphysics \cite{Khairoutdinov2005-fk, Pritchard2014-cv2, Jones2019-ke}. The NNs are trained to emulate subgrid heating ($\Delta T_{\text{phy}}$) in K/s and moistening ($\Delta Q_{\text{phy}}$) in kg/kg/s, where $\Delta T_{\text{phy}}$ represents the combined effect of convective processes, radiation, and subgrid turbulence on heating. This is analogous to a traditional albeit multi-process parameterization. The GCM is a simplified, zonally-symmetric aquaplanet with a full diurnal cycle, fixed season (perpetual austral summer), prescribed sea surface temperatures (SSTs), and a nearly uniformly spaced grid with 64 points along the latitude and 128 points along the longitude dimensions. There are 30 vertical levels within each grid cell, and the GCM has a 30 minute timestep while embedded CRMs have a 20 second timestep. The embedded CRMs have a 4-km horizontal resolution and are each made up of 32 columns. Cloud condensate coupling is omitted from the framework for simplicity. Additional details can be found in the Supplementary Information (SI).

\subsection{Training, Validation, and Offline Test Data}

All NNs were trained on 12 months of longitudinally subsampled simulation data with 30-minute temporal resolution from SPCAM3 and prognostically evaluated for the same time period. Validation data consists of 2 months of simulation data temporally disjoint from the training data. Offline test data consists of 1 week of simulation data with no longitudinal subsampling that is temporally disjoint from both training and validation data. Extending test data volume by a factor of 2 or 4 changes the offline test RMSE by roughly 1 percent, as shown in Figure S28 in the SI. For completeness, we have also included offline results using a multiclimate test set, as shown in Table S2 and Figure S2 in the SI. However, to avoid confusion, we defer discussion of our offline multiclimate results to the Conclusion. Training and validation data retain every $8^{th}$ and $7^{th}$ longitude, respectively. This yields 17,927,168 samples for the training data, 3,556,800 samples for the validation data, and 2,752,512 samples for the offline test data. Training on multiple climates includes one extra sample in the data, but its impact is considered negligible. Inputs are normalized by subtracting the mean and dividing by the standard deviation. For vertically resolved variables, means and standard deviations are calculated at each vertical level. Outputted heating and moistening tendencies are divided by the maximum of the standard deviation and $1 \times 10^{-12}$ at each level. For training, all NNs were trained for a maximum of 180 epochs with a batch size of 4,096 and early stopping set at 5 epochs. Training, validation, and test data were saved to 32 bit precision so that data could fit entirely on GPU RAM, minimizing I/O induced slowdown.

\subsection{End-to-End Pipeline and Analysis}

Our end-to-end pipeline, shown in Figure \ref{fig:pipeline_diagram}, standardizes preprocessing, training, coupling, and analysis across all configurations and is available at \url{https://github.com/SciPritchardLab/ClimScale} \cite{lin_2024_11402897}. Training is parallelized across multiple GPUs using KerasTuner \cite{omalley2019kerastuner}, allowing for the efficient training of hundreds of fully-connected NNs in the span of days using a random search strategy. Coupling to SPCAM 3 is enabled by the Fortran-Keras Bridge (FKB) developed by \citeA{Ott2020-qe}. All hybrid runs are initialized with the same initial condition file used for the reference SPCAM 3 simulation and are each programmed to run for one simulation-year. To mitigate storage demands, every 16th longitude in simulation output is automatically subsampled using GNU parallel \cite{tange_ole_2018_1146014} and NCO \cite{Zender2008-rd}. Better offline performance is characterized by lower offline heating and moistening error on the held-out test set, and better online performance is characterized by lower online temperature and moisture error as well as higher online stability. Online error is only computed for runs that integrate for a full simulation-year without crashing, and hybrid simulations that do not crash within that timeframe are deemed as ``stable." We detect differences in offline and online performance compared to the standard configuration with an array of statistical tests. For comparing average ensemble offline heating and moistening test root mean squared error (RMSE) between configurations, we use two-tailed, two-sample Welch's t-tests (using the \texttt{scipy.stats} library) \cite{2020SciPy-NMeth}. For online ensemble survival rates (i.e., the percentage of hybrid model runs from a configuration that do not crash within one simulation-year) and ensemble-median online temperature and moisture RMSE, we use two-tailed proportion tests and two-tailed permutation tests (using 10,000 permutations), respectively. Our decision to compare ensemble-median RMSE (and not ensemble-mean RMSE) for average ensemble online error is informed by our lack of meaningful online error statistics for crashed runs. Although not formally part of our ablation study, we also test the impact of sources of variation in our search space like dropout (among configurations that vary it), leak (in Leaky ReLu), learning rate, parameter count, choice of optimizer, and batch normalization on offline error, online error, and online stability. Details regarding these tests can be found in Figures S26 through S32 in the SI. Finally, to control for the false discovery rate, we apply a Benjamini-Hochberg correction to the unadjusted p-values before declaring statistical significance at a significance level of 5\% \cite{Benjamini1995-un}. Additional details can be found in the SI while further details regarding ClimScale can be found in Section 6.2.

\subsection{NN Configurations}

To empirically measure the online effects of various design decisions hypothesized to be critical in the literature \cite{Han2020-xy, Han2023-nm, Clark2022-sr, Bhouri2023-pe, Beucler2024-vb, Behrens2024-vj}, we conduct a modified ablation study for a ``standard" configuration that bundles several claims regarding what would be necessary to improve the online behavior of NN convective parameterizations (detailed in Section 2.3.1). In our ablation study, we test two different kinds of modifications, which we will call ``forward" and ``backward" ablations. As with a traditional ablation study in ML, we test the downstream effect of removing a component of the standard (aka baseline) configuration to reveal the importance of the removed component. We call these ``backward ablations". Our ``forward ablations" correspond to configurations that add a feature we suspect could improve online performance. 

Our naive hypothesis is that offline error, online error, and online stability improve together. Correspondingly, we expect the backward ablations to worsen offline and online performance and vice versa.

Since small sample sizes and high variability in online error stemming from stochastic hyperparameter selection can mask differences between configurations, we train 330 NNs with randomly sampled hyperparameters for each of the nine configurations for a total of 2,970 NNs, which is $>27.5\times$ that of \citeA{Ott2020-qe} and at least two orders of magnitude larger than what is traditionally shown in the literature \cite{Rasp2018-fk, Wang2022-po, Iglesias-Suarez2023-qm}. Our hyperparameter search space, shown in Table \ref{tab:searchspace}, is modified from \citeA{Ott2020-qe} to reduce variability resulting from an excessively large, sparse, or suboptimal search space. We found that using a learning rate range that spanned multiple orders of magnitude was a dominant source of variation in validation error, so we narrowed the learning rate range to just one order of magnitude (1e-4 to 1e-3) and sampled it logarithmically. For similar reasons, we decided to sample all possible layer widths within tighter bounds instead of sampling a few possibilities separated by factors of two. Finally, since the Adam optimizer was shown to be dominant in \citeA{Ott2020-qe}, we replaced the stochastic gradient descent (SGD) and RMSprop optimizers with Rectified Adam and Quasi-Hyperbolic Adam, which performed similarly to Adam in preliminary experiments and have been shown to be performant in other contexts \cite{Ma2018-lv, Liu2019-hk, Geleta2023-dg}. Put together, these interventions result in distributions of online temperature and moisture error that are concentrated within one order of magnitude (instead of two as in \citeA{Ott2020-qe}).

\begin{table}
\centering
\begin{tabular}{l l}
\hline
  Hyperparameter  & Range  \\
\hline
   Hidden layers  & $\llbracket $ 4, 11 $\rrbracket $  \\
   Nodes per layer  & $\llbracket $ 200, 480 $\rrbracket $  \\
   Batch normalization  & \{On, Off\}   \\
   Dropout  & [0.0, 0.25]   \\
   Optimizer  & \{Adam, RAdam, QHAdam\}  \\
   Leaky ReLu slope  & [0.0, 0.4]   \\
   Learning rate  & [1e-4, 1e-3]  \\
\hline
\end{tabular}
\caption{Common NN hyperparameter search space shared by all tested parameterization design configurations (excluding the no dropout configuration). All NNs tested here are fully-connected NNs. RAdam and QHAdam stand for ``Rectified Adam" and ``Quasi-hyperbolic Adam", respectively. Each hyperparameter is sampled uniformly at random, logarithmically in the case of learning rates. \label{tab:searchspace}}
\end{table}

All nine configurations share a set of inputs similar to those used by \citeA{Rasp2018-fk} and \citeA{Ott2020-qe}: vertically resolved coarse-scale temperature and humidity variables as well as scalars for surface pressure, top-of-atmosphere insolation, surface sensible heat flux, and surface latent heat flux. Output variables across all configurations are identical, consisting of vertically resolved heating and moistening tendencies. In all cases, the top five vertical levels (corresponding to anything approximately above 55 hPa) of moisture and moistening are not included during training and set to zero during coupling, in line with findings from \citeA{Clark2022-sr} and \citeA{Brenowitz2020-bv} that negligible magnitudes and causatively irrelevant variability at these altitudes can inhibit the machine learning. NNs are not trained to directly output radiative fluxes and precipitation as those variables are diagnostic (uncoupled) in an aquaplanet setting (i.e., without land). The complete list of inputs and outputs for the standard configuration is shown in Table \ref{tab:inputsoutputs}.

\begin{table}[h]
\centering
\begin{tabular}{l c c l c c}
\hline
\multicolumn{3}{l}{\textbf{Input variables}} & \multicolumn{3}{l}{\textbf{Output variables}} \\
\hline
\textbf{Variable} & \textbf{Unit} & $N_z$ & \textbf{Variable} & \textbf{Unit} & $N_z$  \\
\hline
Temperature & K & 30 & Heating rate $\Delta T_{\text{phy}}$ & K s$^{-1}$ & 30 \\
Relative Humidity* & \% & 25 & Moistening rate $\Delta Q_{\text{phy}}$ & kg kg$^{-1}$ s$^{-1}$ & 25 \\
$(t-1)$ Heating rate $\Delta T_{\text{phy}}$* & K s$^{-1}$ & 30 &  &  &  \\
$(t-1)$ Moistening rate $\Delta Q_{\text{phy}}$* & kg kg$^{-1}$ s$^{-1}$ & 25 &  &  &  \\
Surface pressure & Pa & 1 &  &  &  \\
Incoming solar radiation & W m$^{-2}$ & 1 &  &  &  \\
Sensible heat flux & W m$^{-2}$ & 1 &  &  &  \\
Latent heat flux & W m$^{-2}$ & 1 &  &  &  \\
Meridional wind* & m s$^{-1}$ & 30 &  &  &   \\
Ozone volume mixing ratio* & m$^{3}$ m$^{-3}$  & 30 &  &  &   \\
Cosine of zenith angle* &  & 1 &  &  &   \\
\hline
\multicolumn{2}{l}{Size of stacked vectors} & 175 & \multicolumn{2}{l}{} & 55 \\
\hline
\end{tabular}
\caption{Table showing input and output variables, their units, and their number of vertical levels $N_z$ for the standard configuration. $(t-1)$ refers to the value corresponding to the previous timestep. Variables indicated with an asterisk are removed or transformed in configurations corresponding to backward ablations. Stratospheric moisture (corresponding to the top five levels) is nearly always zero and excluded from training. \label{tab:inputsoutputs}}
\end{table}

\subsubsection{Standard Configuration [Baseline]}

The \textbf{standard} configuration is our baseline against which other configurations are compared. Although there is some design similarity to early work of \citeA{Rasp2018-fk}, additional heuristics like the use of a relative humidity transformation for moisture input \cite{Beucler2024-vb}, removing stratospheric moisture and moistening \cite{Brenowitz2020-bv, Clark2022-sr}, including convective memory \cite{Han2020-xy, Han2023-nm}, expanding the input vector to include the mixing ratio of the volume of ozone and the cosine of the zenith angle, and normalizing the outputs by their standard deviation (per vertical level) are also included to provide for a skillful preliminary baseline. Several of these heuristics are empirically tested in the other configurations listed below. The complete list of inputs and outputs for this configuration is shown in Table \ref{tab:inputsoutputs}.

\subsubsection{Specific Humidity Configuration [Backward ablation]}
The \textbf{specific humidity} configuration reverses the relative humidity transformation for moisture input in the standard configuration in order to see if the offline benefits of a ``climate-invariant" transformation discovered in \citeA{Beucler2024-vb} also manifest in improved performance online. \citeA{Beucler2024-vb} identified multiple feature transformations that improve generalization across climates in an offline setting by decreasing the number of situations in which NNs extrapolate out-of-distribution. In the context of moisture, the marginal distribution of relative humidity changes very little in warmer climates (when measuring change using Hellinger distance) and is bounded by design, except in cases of supersaturation \cite{Beucler2024-vb}. While the NNs are tested online in the same climate, the implicit hypothesis is that online errors may include pathologies in which the ML-coupled fluid dynamics lead the input state vector out-of-distribution from the training set, such that climate-invariant feature transformations may be beneficial. Consequently, we expect a configuration that removes such a transformation to worsen offline and online performance relative to our baseline.

\subsubsection{No Memory Configuration [Backward ablation]}
The \textbf{no memory} configuration omits convective memory (i.e., heating and moistening tendencies from the previous timestep) from the input. Including such memory was argued to be beneficial online in \citeA{Han2023-nm} and offline in \citeA{Han2020-xy} while similar benefits were shown using precipitation from a previous timestep in \citeA{Behrens2024-vj}. In other work, \citeA{Colin2019-ks} and \citeA{Shamekh2023-ed} formally reveal the existence of a persistent ``microstate memory'' due to convective organization in the CRMs that our NNs with memory are trained to emulate. Thus we expect the configuration without memory to worsen offline and online performance relative to our baseline.

\subsubsection{No Wind Configuration [Backward ablation]}
The \textbf{no wind} configuration excludes meridional wind speeds from the input, which was included in previous work from \citeA{Rasp2018-fk} and \citeA{Ott2020-qe} but argued to be of secondary importance by \citeA{Han2020-xy} and  \citeA{Mooers2021-sh}. Since this dimension of the horizontal wind field theoretically has the capacity to constrain the convection in SPCAM due to organizing effects of wind shear, we expect this configuration to worsen offline and online performance relative to our baseline.

\subsubsection{No Ozone Configuration [Backward ablation]}
The \textbf{no ozone} configuration omits ozone mixing ratio from the input vector similar to previous work \cite{Rasp2018-fk, Ott2020-qe}. Because ozone is causally relevant to radiative heating in the stratosphere and is prescribed as a function of latitude in the climate simulator, its omission in \citeA{Ott2020-qe} and \citeA{Rasp2018-fk} was a mistake. Consequently, we expect this configuration to worsen offline and online performance relative to our baseline.

\subsubsection{No Zenith Angle Configuration [Backward ablation]}
The \textbf{no zenith angle} configuration omits the cosine of the zenith angle from the input vector similar to previous work \cite{Rasp2018-fk, Ott2020-qe}. Calculating optical depth is a crucial part of radiative transfer models. Given that the zenith angle is a key factor in optical depth calculations, we expect omitting it to worsen online and offline performance. 

\subsubsection{Mean Absolute Error (MAE) Configuration [Forward ablation]}
The \textbf{MAE} configuration swaps out the Mean Squared Error (MSE) loss used for training in all other configurations with a Mean Absolute Error (MAE) loss. Compared to the MSE loss, the MAE loss is designed to learn the conditional median rather than the conditional mean, making it potentially less sensitive to outliers in the data as well as asymmetric distributions. In our context, our experimentation with the MAE loss is primarily motivated by the fact that the marginal distributions of different tendency variables appear to be more Laplacian than Gaussian. For Laplacian-distributed errors, an MAE loss is more appropriate \cite{Hodson2022-fm}. Although we do not have access to the conditional distributions of the outputs since we do not have multiple realizations of the CRM given a single input, we suspect that they are similar to the marginal distributions. If this is the case, we expect a MAE loss to improve offline and online performance. The marginal distributions of the heating and moistening tendencies are shown in Figures S34 and S35 in the SI.

\subsubsection{No Dropout Configuration [Forward ablation]}
The \textbf{no dropout} configuration removes dropout from the search space shown in Table \ref{tab:searchspace}. In many ML contexts, dropout is typically used as a regularization tool to prevent overfitting \cite{Hinton2012-hi, Srivastava2014-cy, Molina2021-on} but also has seen application as a computationally-efficient tool for representing model uncertainty \cite{Gal2015-ss, Behrens2024-vj}. While \citeA{Ott2020-qe} did not find a strong relationship between dropout and online performance, \citeA{Behrens2024-vj} found that using even a modest amount of dropout (as small as .05) in the last hidden layer of a NN resulted in noticeable deterioration in offline skill. Given recent experience from \citeA{Behrens2024-vj}, we expect eliminating dropout completely to improve both offline and online performance.

\subsubsection{Multiclimate Configuration [Forward ablation]}
The \textbf{multiclimate} configuration is trained on three different climates, one of which is identical to that used for other configurations. The other two are created by prescribing -4K colder and +4K warmer SSTs and waiting for the atmosphere to equilibrate. The number of training data samples used per climate is roughly ~1/3 of the total number of training data samples used in the other configurations. Training on multiple climates is argued as helpful to reduce extrapolation error in prior work \cite{Clark2022-sr, Bhouri2023-pe, Lin2024-zo}, and, in general, diversifying training data is a well-known and intuitive heuristic for improving the generalization performance of deep learning models. Intuitively, we expect training on multiple climates to improve offline and online performance.

\section{Results (in-distribution)}

\subsection{Offline Results}

\begin{figure}[!htbp]
 \centering
 \includegraphics[width=\textwidth]{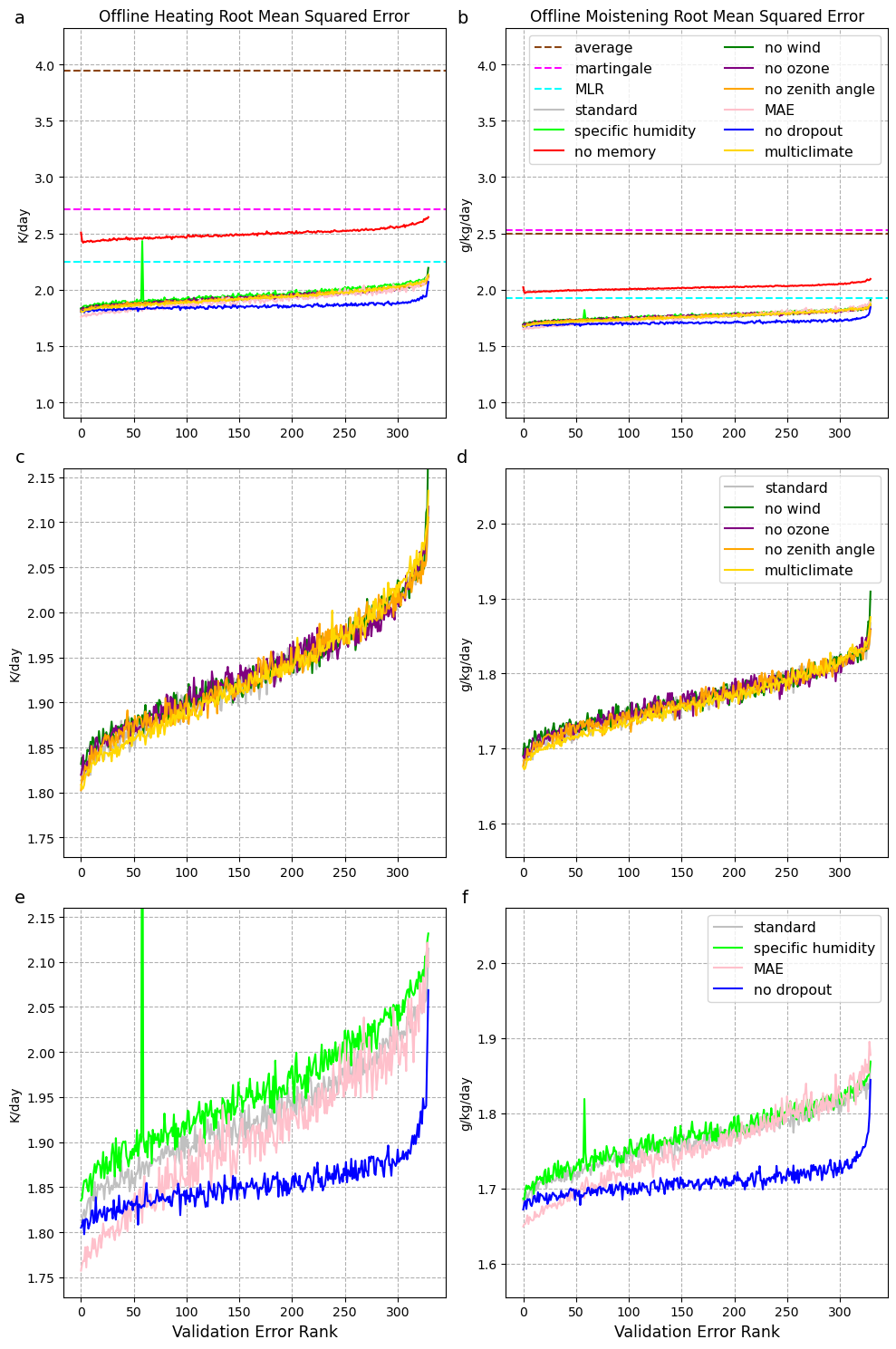}
 \setlength{\belowcaptionskip}{-1em}%

\caption{Offline test RMSE for subgrid heating (1a,1c,1e) and moistening (1b,1d,1f) tendencies across configurations are plotted against validation error rank. 1a and 1b show RMSE for average, multiple linear regression (MLR), and martingale baselines. The MLR baseline makes use of inputs and outputs from the standard configuration. 1e and 1f show configurations with statistically distinct average RMSE, and 1c and 1d show the others. Figure S2 in the SI presents an analogous figure using a multiclimate test set.}
 \label{fig:Offline_Error_Complete}
\end{figure}

To avoid confusion, we present our offline results in this section within the context of \textit{in-distribution} offline error. Although differences emerge when using an out-of-distribution (i.e. multiclimate) offline test set, we defer the discussion of out-of-distribution differences to Section 5.1 in the Conclusion. Figure \ref{fig:Offline_Error_Complete} tests our expectations for in-distribution offline error. The comprehensive sampling (330 NNs per configuration) shows offline test error values ranked by validation error for each configuration. We measure the offline effects of each backward and forward ablation via two-tailed Welch's t-tests on the differences in ensemble-average in-distribution offline heating and moistening errors for each ablation when compared to the standard configuration. For reference, we also show the offline test errors for a simple average, a multiple linear regression (MLR), and using the previous tendency values to predict the following one (i.e. making use of a martingale assumption).

Several of our expectations are confirmed. Among configurations corresponding to backward ablations, removing memory stands out as having the largest impact as it increases the ensemble-average offline RMSE substantially by 0.567 K/day (0.256 g/kg/day) for mean heating (moistening) (Figure \ref{fig:Offline_Error_Complete} a,b). The offline impact of reformulating the moisture input from relative to specific humidity is less extreme yet nonetheless detectable, increasing ensemble-average offline RMSE by .0288 K/day (.00789 g/kg/day). In terms of forward ablations, the MAE and no dropout configurations both show offline improvement, decreasing ensemble-average offline RMSE by .0226 K/day (.00952 g/kg/day) and .0765 K/day (.0545 g/kg/day), respectively. These differences are all statistically significant.

Other expectations are invalidated. Our backward ablations of removing wind, ozone, and zenith angle from the input vector did not statistically affect offline error (Figure \ref{fig:Offline_Error_Complete} c,d). Our forward ablation of training on multiple climates also had no statistically detectable effect. Since \citeA{Ott2020-qe} and \citeA{Wang2022-po} both showed a relationship between offline error and online performance (albeit solely in the form of online simulation stability), we might expect inter-configuration differences in online error (and online stability) to mirror differences identified offline. However, these relationships, especially with respect to offline and online error specifically, have not been statistically evaluated in previous works (e.g., \citeA{Ott2020-qe, Wang2022-po}).

If \textit{in-distribution} offline error is predictive of \textit{in-distribution} online performance in general, Figure \ref{fig:Offline_Error_Complete} implies we should expect both online stability and online error to be worse for the no memory and specific configurations, statistically the same for the no wind, no ozone, no zenith angle, and multiclimate configurations, and better for the MAE and no dropout configurations. We explore this topic in the next section.

\subsection{Online Results}

\begin{table}[h]
\small
\centering
\begin{tabular}{l c c c c c c c}
\hline
\textbf{Configuration} & \textbf{Survival} & \textbf{Online Error (K)} & \textbf{Offline Error (K/day)} & \textbf{Spearman's $\rho$}\\
\hline
standard          & 42.7\% & 3.48 K & 1.93 K/day & \textbf{.855} \\

specific humidity & \textcolor{red}{\textbf{-26.1\%}} & +.130 K & \textcolor{red}{\textbf{+.0288 K/day}} & \textbf{.834} \\

no memory         & \textcolor{red}{\textbf{-32.7\%}} & \textcolor{red}{\textbf{+.800 K}} & \textcolor{red}{\textbf{+.567 K/day}} & .311 \\

no wind           & +1.82\% & -.00430 K & +.00526 K/day & \textbf{.789} \\

no ozone          & \textcolor{teal}{\textbf{+13.3\%}} & \textcolor{red}{\textbf{+.212 K}} & +.00443 K/day & \textbf{.803} \\

no zenith angle   & \textcolor{teal}{\textbf{+11.2\%}} & -.142 K & +.00226 K/day & \textbf{.829} \\

MAE               & \textcolor{teal}{\textbf{+32.4\%}} & \textcolor{red}{\textbf{+1.16 K}} & \textcolor{teal}{\textbf{-.0226 K/day}} & \textbf{.706} \\

no dropout        & \textcolor{red}{\textbf{-14.2\%}} & \textcolor{teal}{\textbf{-.967 K}} & \textcolor{teal}{\textbf{-.0765 K/day}} & \textbf{.353} \\

multiclimate      & \textcolor{teal}{\textbf{+27.3\%}} & \textcolor{teal}{\textbf{-.237 K}} & -.000903  K/day & \textbf{.858} \\
\hline

\end{tabular}
\caption{The columns titled Survival, Online Error (K), Offline Error (K/day), and Spearman's $\rho$ in Table 3 refer to online survival rate (i.e., percentage of runs that integrated the full simulation year without crashing), ensemble-median online temperature RMSE in K, ensemble-average offline heating RMSE in K/day, and Spearman correlation between offline heating and online temperature RMSE. Online error statistics and offline-online relationships exclude runs that crash. Excluding the column showing correlations, all rows following the first denote anomalies relative to the standard configuration's statistics. Significant anomaly values are bolded, and color indicates direction of difference (with red indicating worse and teal indicating better performance). Statistics are deemed significant at $\alpha = 0.05$ after applying a Benjamini-Hochberg correction, and Figure S1 in the SI shows this correction being applied.  The analogous table for online moisture and online moistening error is similar and can be found in Table S1 in the SI. Table S2 in the SI shows similar statistics but on a multiclimate test set.}
\label{tab:summarytable}
\end{table}

\begin{figure}[!htbp]
 \centering
 \includegraphics[width=\textwidth]{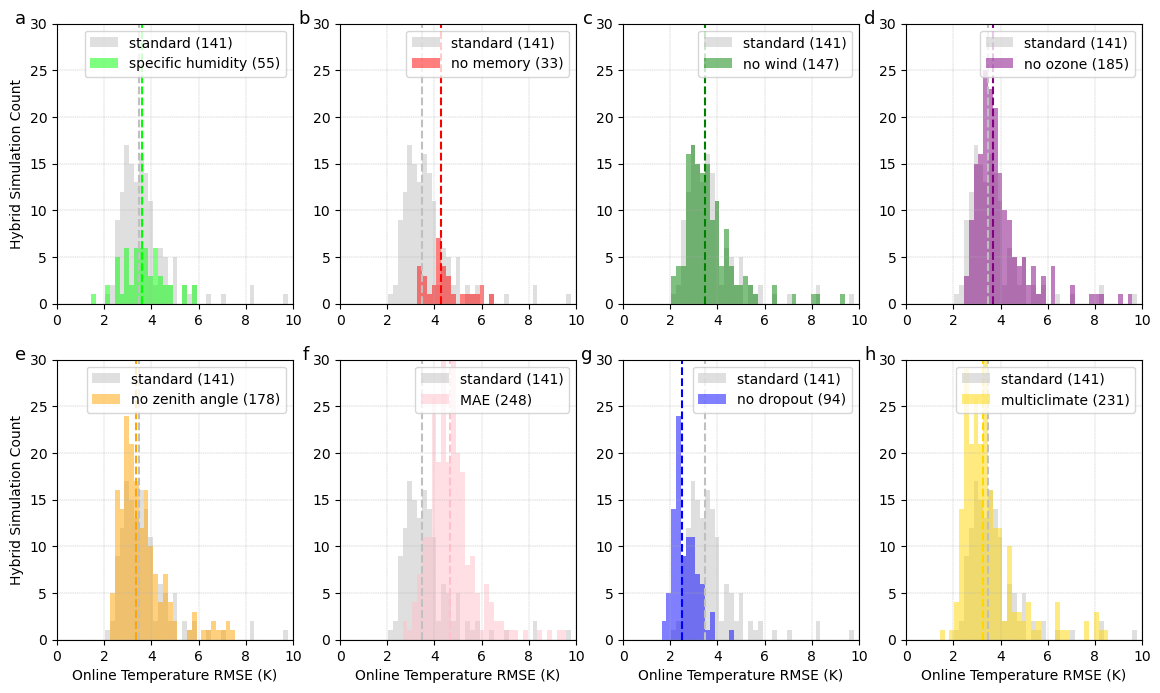}
 \setlength{\belowcaptionskip}{-1em}%

\caption{Histograms of online temperature RMSE in K for each configuration compared against that of the standard configuration are shown here. Only models that integrated for the entire simulation duration (i.e., did not crash while integrating a full simulation year) are shown, and number of surviving models per configuration is shown in the legend. Vertical lines correspond to ensemble-median online temperature RMSE. Bin-width $= \frac{10}{49} K \approx .204$ K.}
 \label{fig:online_temperature_error}
\end{figure}

\begin{figure}[!htbp]
 \centering
 \includegraphics[width=\textwidth]{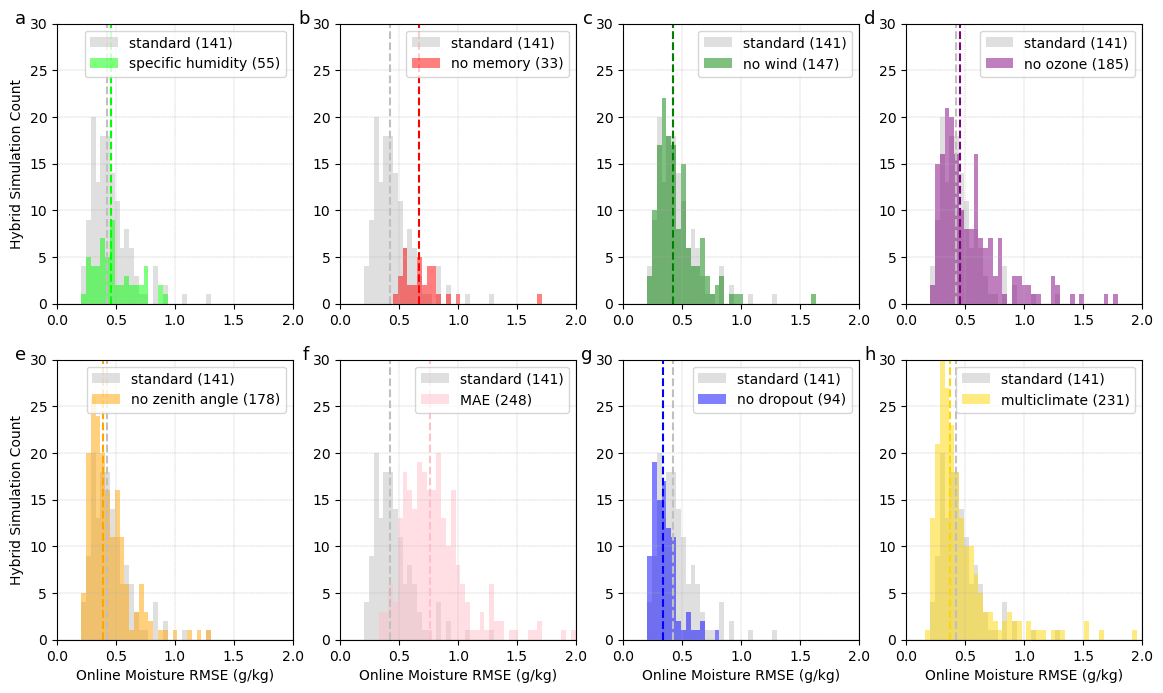}
 \setlength{\belowcaptionskip}{-1em}%

\caption{Histograms of online moisture RMSE in g/kg for each configuration compared against that of the standard configuration are shown here. Only models that integrated for the entire simulation duration (i.e., did not crash while integrating a full simulation year) are shown, and number of surviving models per configuration is shown in the legend. Vertical lines correspond to ensemble-median online moisture RMSE. Bin-width $= \frac{2}{49} \approx .0408$ g/kg.}
 \label{fig:online_moisture_error}
\end{figure}

We present online survival statistics, offline heating error, online temperature error, and Spearman rank correlations between offline heating and online error for each configuration in Table \ref{tab:summarytable} (analogous statistics for offline moistening error and online moisture error can be found in Table S2 in the SI). Histograms of online temperature and moisture RMSE for simulations that do not crash can be found in Figure \ref{fig:online_temperature_error} and Figure \ref{fig:online_moisture_error}, respectively. Our results immediately invalidate our naive hypothesis that offline error, online error, and online stability improve together. More specifically, we can see that the ``MAE" and ``no dropout" forward ablations have divergent effects on online error and online stability despite both reducing offline error. Using an MAE, instead of MSE, loss results in the highest improvement in online stability but the worst deterioration in online error. At the same time, removing dropout grants the highest improvement to online error but comes at the cost of reduced online stability. In addition to disproving our naive hypothesis, this contradicts prior practice of relying on stability as a proxy for general online performance and disproves the notion that improving performance across one online metric implies simultaneous improvement on all others \cite{Ott2020-qe, Gagne2020-av, Wang2022-po}.

Other results are less surprising. The backward ablation of excluding memory unambiguously worsens offline and online performance across all metrics: offline error, online stability, and online error. Conversely, although the multiclimate configuration had no detectable offline impact, it is the only ablation to statistically improve upon the standard configuration on every online metric, as seen in Figures \ref{fig:online_temperature_error}h and \ref{fig:online_moisture_error}h. The no wind, no ozone, and no zenith angle configurations also had no detectable offline impact, but, the no ozone configuration does have statistically higher ensemble-median online temperature RMSE, a reassuring expected finding. Despite the importance of zenith angle in calculating optical depth, we suspect it had less of an impact on the ensemble-median temperature RMSE because of a high R-squared between cosine of zenith angle and solar insolation ($> .99$ when values corresponding to zero solar insolation are masked out). Surprisingly, the no ozone and no zenith angle configurations both have improved stability. We speculate this is because certain input features are redundant and causal pruning may be necessary to avoid learning spurious correlations that fail to generalize outside the training data \cite{iglesias2024causally}. As for the remaining backward ablation, the specific humidity configuration worsens stability but does not have a statistically detectable impact on online error. In Section 4, we provide further details regarding what sample size we believe may be necessary for robustly detecting causally relevant differences online. 

When it comes to the relationship between offline and online error, rank correlations in Table \ref{tab:summarytable} and Table S1 indicate that such a relationship exists for configurations with convective memory and non-zero dropout. However, when eliminating dropout entirely, the offline-online error relationship is weakened, but still significant, for offline heating and online temperature and no longer present for offline moistening and online moisture. For configurations that vary it, dropout is a dominant source of variation for both offline and online error, as seen in Figure S26. Other aspects of our search space also have notable impacts on offline error, online error, and online stability, shown in Figures S27 to S32. In particular, batch normalization presents benefits on all fronts, as seen in Figure S31 and Figure S32f. In Figures S28, S29, and S32c we also see some evidence that smaller learning rates improve offline heating error, online temperature error, and online stability while higher parameter counts reduce offline heating error, offline moistening error, and online temperature error. Finally, we find that the inclusion of the QHAdam optimizer in our search space was a mistake, as it has a detrimental effect on online stability when compared to the alternatives, as seen in Figure S32e.

To ensure the robustness of our conclusions, we investigate the possibility that our decision to exclude crashed simulations in our comparisons of online error may have biased our results. For the ablation study, this could happen in cases where simulations with high online error crash less frequently in one configuration than another, resulting in a higher ensemble-median online error than would have otherwise been recorded. Under such circumstances, we would expect a pairwise comparison of ensemble-median online errors for the standard configuration and some ablation to be different at the end of the simulation-year than at some point during the simulation where the high online errors of the (not yet) crashed simulations are still being recorded. However, as seen in Figure S13, intersections of the standard configuration's ensemble-median online error growth only occur with those of ablations whose online error we already consider to be statistically identical. We believe this is sufficient evidence to conclude that this potential for bias has not impacted the results of our ablation study. We conduct a similar check for our finding that batch normalization reduces online temperature and moisture error in Figure S33, and conclude that that result is also not impacted. Readers interested in more detail can find additional figures in the SI showing time-series plots of online error growth (S4 to S12), offline and online error scatterplots (S16 to S24), and histograms for the number of months integrated for each configuration (S15).

\subsection{Persistent Online Zonal Mean Biases}

\begin{figure}[!htbp]
 \centering
 \includegraphics[width=\textwidth]{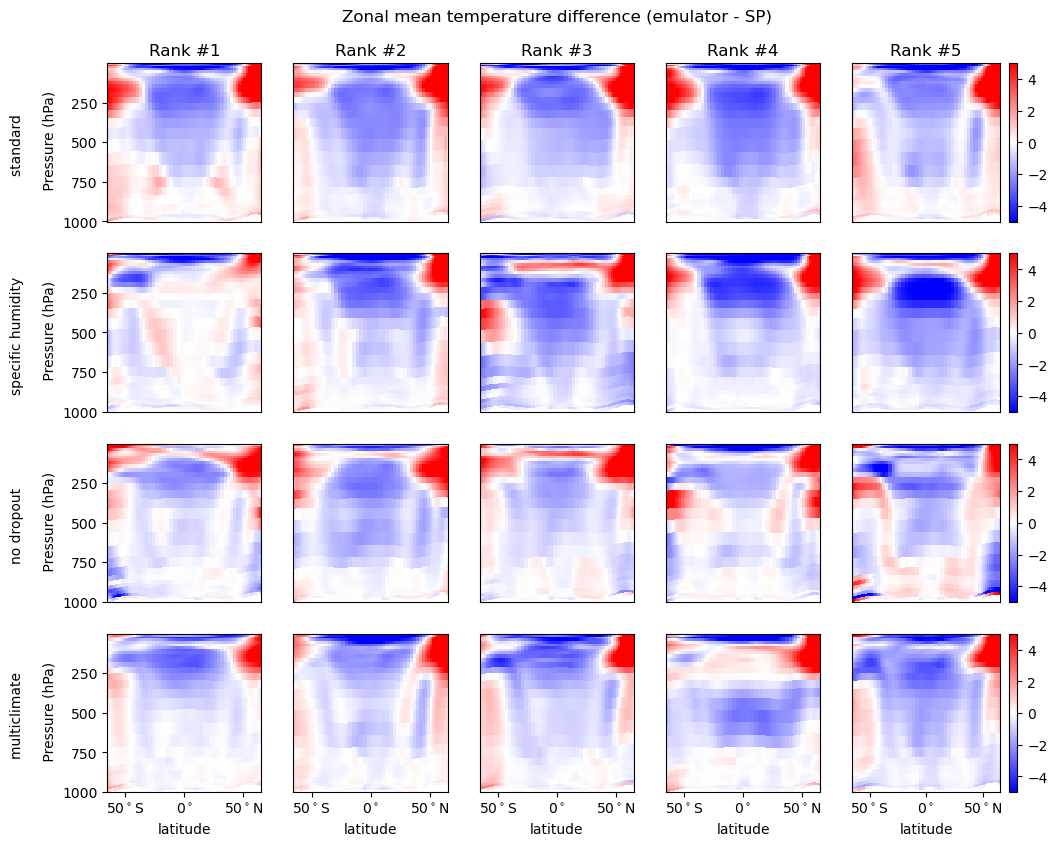}
 \setlength{\belowcaptionskip}{-1em}%

\caption{This figure shows zonal mean temperature biases for the top five hybrid models across the standard, specific, no dropout, and multiclimate configurations. Models to the left have lower online error.}
 \label{fig:zonal_bias}
\end{figure}

Our online results show the effect of various ablations on aggregate statistics across $\mathcal{O}(100)$ size ensembles. However, a closer look at the best online simulations across different configurations reveals common patterns in zonal mean biases relative to the reference SPCAM simulation, with a cold bias in the stratosphere and a warm bias in the poles (as seen in Figure \ref{fig:zonal_bias}). Systematic biases in the stratosphere and near the poles have also been reported by others \cite{Wang2022-po, iglesias2024causally}, and our results confirm that the biases experienced in our work are reproducible. However, we also show that certain interventions can change the structure of this bias. For example, training on multiple climates results in an improved, asymmetric zonal mean bias, indicating that while persistent, this bias structure can be systematically mitigated. Nevertheless, our bias structure is very sensitive to hyperparameter sampling, as the hybrid simulation with the smallest zonal biases comes from the specific humidity configuration, which is less stable than the standard configuration and consistently has higher offline temperature and moisture RMSE than the no dropout configuration. Sensitivity to hyperparameter tuning and high downstream, online noise may imply that reliably mitigating this bias structure in future work will also require testing additional hypotheses with large ensembles.

\section{Estimating necessary sample size for robust detection}

While the scale of our online sampling is extensive by design, one might naturally wonder whether or not this level of scale is actually necessary. Could our results be reproduced using a smaller sample size and fewer computational resources? We chose a large sample size of 330 NNs because the dispersion of online error seen in previous work like that of \citeA{Ott2020-qe} spanned two orders of magnitude. However, as seen in Figures \ref{fig:online_temperature_error} and \ref{fig:online_moisture_error},  our online errors only span one. In this section, we estimate the sample size necessary to robustly detect meaningful differences online.

Selecting a sample size for non-parametric tests---especially for statistics like medians which do not obey the usual Central Limit Theorem---is not firmly established; there exist multiple methods but they are imperfect and make use of simplifying assumptions \cite{Noether1987-po, Hamilton1991-xa}. The situation is further complicated by the fact that many hybrid runs may crash, and only those completing the entire simulation are used for online error statistics.

To establish a conservative estimate, we make use of our least detectable (or highest significant p-value) yet scientifically justified finding---that not including ozone in the input increases temperature error online. In line with common practice, we choose a significance level, $\alpha$, of $5\%$ and a power, $1 - \beta$, of $80\%$. We also make several simplifying assumptions.

\begin{itemize}
\item An equal proportion, $\gamma$, of NNs integrate without crashing in both configurations. More specifically, this proportion is equal to the survival rate of the standard configuration.
\item If the distributions of online error for both configurations vary, they only vary in location, not shape, skew, or dispersion (i.e., the distribution of online error for one configuration can be approximated by shifting the other).
\end{itemize}

Under these assumptions, the Mann Whitney U-Test (otherwise known as the Wilcoxon ranked sum test) is equivalent to a test of a difference in medians and provides a closed form for power calculations \cite{Noether1987-po, Hamilton1991-xa}. In this case, we apply the Mann Whitney U-Test to the online RMSEs for the standard configuration and a version of those same online RMSEs offset by the difference in ensemble-median RMSE when compared to the no ozone configuration. We arrive at a sample size estimate of 384, demonstrating the value of $\mathcal{O}(100)$ ensembles for detecting causally relevant differences in online error. We hope this sample size estimate serves as a useful reference point, and not a canonical requirement, for future work. Larger effect sizes require far fewer samples for detection. As an example, detecting the shifts in ensemble-median online temperature error for the multiclimate and no dropout configurations would require sample sizes of 312 and 29, respectively. Making use of a baseline with improved stability and a search space with less variation may also reduce the sample size necessary to detect a given effect size. Further details regarding the sample size estimation can be found in Text S3 in the SI.

\section{Conclusion}

Leveraging an end-to-end pipeline for $\mathcal{O}$(100) ensembles of hybrid physics-ML simulations, we arrive at empirically robust recommendations for the development of future NN parameterizations of moist convection and identify counterintuitive nuances regarding the relationships between offline error, online error, and online stability. We summarize our findings below, categorized by primary and secondary importance:

\underline{Primary Findings:}
\begin{itemize}
    \item Using memory (to implicitly represent convective memory), batch normalization, and training on multiple climates reduce offline and online error and improve online stability.
    \item Dropout negatively impacts both offline and online error, but setting the dropout rate to zero reduces stability.
    \item Switching from an MSE to an MAE loss also improves offline error, but online error is worsened while online stability is improved.
    \item Using a ``climate-invariant" feature transformation from specific to relative humidity and removing redundant input variables (i.e. feature pruning) both improve stability.
    \item Detecting causally relevant differences in online statistics with 80\% power may require $\mathcal{O}$(100) ensembles of hybrid simulations.
\end{itemize}

\underline{Secondary Findings:}
\begin{itemize}
    \item Smaller learning rates are weakly associated with lower offline heating error, lower online temperature error, and higher stability.
    \item Higher parameter counts are weakly associated with lower offline heating and moistening error as well as lower online temperature error.
    \item Use of the QHAdam optimizer (as configured in our training code) decreases online stability compared to Adam and RAdam. 
\end{itemize}

When reporting relationships between offline error, online error, and online stability, we use an in-distribution offline test set to avoid confusion and maintain narrative consistency. However, if the root cause of online instability and error is a failure to generalize out-of-distribution, the use of an out-of-distribution offline test set may result in a more reliable offline proxy for online error. To check whether or not this is the case, we also show rank correlations between multiclimate offline error and ``in-distribution" online error in Table S2 in the SI. We find that these rank correlations are actually weaker than those for in-distribution offline and online error (with the exception of the multiclimate configuration, as seen in Table \ref{tab:summarytable} and Table S1). That being said, there are a few differences worth noting, e.g. the multiclimate offline error is significantly lower with the multiclimate configuration and higher with the specific humidity configuration, validating results from \citeA{Beucler2024-vb}. Additionally, the offline moistening error for the no dropout configuration is higher, not lower, than that of the standard configuration with a multiclimate test set. Nevertheless, it is possible that there exists a more reliable offline proxy for online error than the ones evaluated in this paper. Such a proxy would enable faster progress in NN parameterization development by augmenting the utility of more modern optimizers and optimization algorithms. Additionally, lessons learned from open competitions that crowd-source offline error optimization, like the recent 2024 LEAP Atmospheric Physics using AI (ClimSim) Kaggle competition, could become more relevant to the online hybrid climate modeling evaluation task \cite{leap-atmospheric-physics-ai-climsim}.

Limitations in our work may qualify the claims we are able to make, but they also point to promising and concrete directions for future research. While our work is empirical, dynamical analyses in follow-up work may unveil the root causes for why using an MAE instead of MSE loss and removing dropout have opposite effects on online error and online stability (despite both reducing offline error). They may also identify physically-informed ways to systematically eliminate the online stratospheric zonal mean biases seen in this and prior work. Such advances may result in scenarios in which high online skill and stability are the norm and not the exception, warranting longer, multi-year integration runs and the use of a more complex GCM (as opposed to a simplified, zonally-symmetric aquaplanet with prescribed SSTs). Given how the online temperature and moisture RMSEs often continue to increase for many hybrid simulations that are deemed ``stable" by virtue of having completed one year of integration without crashing (as seen in Figures S4 through S13), multi-year integrations in future work may also be necessary to verify whether or not one-year integrations are a valid proxy for stability in general. For the purposes of this study, we believe one-year integrations may be sufficient since $>50\%$ of hybrid runs crash within one simulation-year in our baseline and, in general, simulations that do crash most frequently crash within the first month, as seen in Figure S15. Nevertheless, true step-change improvements in both online error and online stability for closer-to-operational settings will likely require moving beyond fully-connected NNs to more sophisticated architectures, as showcased by \citeA{Hu2024-os}. Even further advancements may require the use of coupled ensembles, dynamics-informed curation of training data, additional ``climate-invariant" feature transformations, and inherently stochastic models that better represent the coarse-grained effects of convection and turbulent flow \cite{Behrens2024-vj, Balogh2021-kw, Beucler2024-vb, Yu2023-on, Kohl2023-dr}.

Nevertheless, our results demonstrate that large-scale sampling is a vital tool for navigating noise and nuance in the online behavior of ML parameterizations. The dispersion in online RMSE or survival rates from hyperparameter tuning can be large enough to obscure the impact of different design choices, and it is possible to train a superior model (i.e., when evaluated online) using seemingly suboptimal design choices offline, as evidenced by a single NN from the specific humidity configuration having the lowest online zonal mean temperature biases across all 2,970 models evaluated (shown in Figure \ref{fig:online_temperature_error}a). Had we drawn conclusions from this singular, performant NN parameterization, we would have come to a conclusion contrary to that implied by the ensemble at large. Although hundreds of ensemble members may be necessary to detect small yet statistically significant and causally relevant differences, our work suggests that, at a minimum, ensemble sizes of a few dozen should be used to detect larger effect sizes with sufficient power. While there exist multiple sophisticated strategies for mitigating emergent effects of online behavior like online coupled learning as described in \citeA{Rasp2020-ns}, gradient-free ensemble Kalman methods as in \citeA{Lopez-Gomez2022-sy}, and creating a fully differentiable hybrid physics-ML atmospheric model as demonstrated by \citeA{Kochkov2023-bd} with NeuralGCM, sampling at sufficient scale may ultimately still be necessary to empirically settle competing hypotheses and make reproducible progress.

Beyond climate simulation, the lessons learned in this paper are relevant to hybrid (ML-physics) simulations across the geosciences. We have demonstrated a standardized process for distinguishing emergent signal from noise in the online behavior of ML parameterizations. Our view is that without infrastructure for such rapid, large-scale downstream online tests, development of hybrid ML-physics models can be muddled. This arises from the difficulty of comparing configurations of ML-parameterizations whose distributions of online error fail to converge due to insufficient sampling and the technical complexity of sampling at sufficient scale. Evidently, the climate model use case is one in which high dispersion owing to stochastic aspects of ML optimization, such as architecture searches, can mask modest but real gains in downstream performance. It can also give a false impression of the benefits of a given design decision. This can ultimately manifest in premature conclusions and missed opportunities. We hope our work motivates a transition from extrapolating from anecdotal experiences or small ensemble sizes to adopting a more systematic and reproducible approach to ML parameterization development.

%
%

\section*{Open Research}


Version v2 of spcam3.0-neural-net-spreadtesting, the codebase used for creating the reference SPCAM3 simulation and dynamical core for online simulations is preserved at
\url{https://zenodo.org/records/11392025}, available via Apache License 2.0 and developed openly at \url{https://github.com/SciPritchardLab/spcam3.0-neural-net} \cite{Pritchard2024-ac}. v4 of ClimScale used for conducting hybrid physics-ML climate runs is preserved at \url{https://zenodo.org/records/14252488}, available via Apache License 2.0 and developed openly at \url{https://github.com/SciPritchardLab/ClimScale} \cite{lin_2024_11402897}. All experiments were run on NVIDIA V100 GPUs. Approximately 2,290 GPU-hours were used to train 2,970 NNs across all nine configurations.

\acknowledgments
This work is primarily funded by National Science Foundation (NSF) Science and Technology Center (STC) Learning the Earth with Artificial Intelligence and Physics (LEAP), Award \# 2019625-STC. High-performance computing was facilitated by Bridges\-2 at the Pittsburgh Supercomputing Center (PSC) through allocation ATM190002 from the Advanced Cyberinfrastructure Coordination Ecosystem: Services \& Support (ACCESS) program, which is supported by NSF grants \#2138259, \#2138286, \#2138307, \#2137603, and \#2138296. MP acknowledges co-funding from the US Department of Energy (DE-SC0023368). Tom Beucler acknowledges partial funding from the Swiss State Secretariat for Education, Research and Innovation (SERI) for the Horizon Europe project AI4PEX (Grant agreement ID: 101137682). Eliot Wong-Toi is funded by the Hasso Plattner Research School at UC Irvine. We are thankful to the system administrators at PSC, in particular Tom Maiden and TJ Olesky, as well as David Walling for his assistance with HPC support through the XSEDE Extended Collaborative Support Service program. We extend our gratitude to Dave Lawrence (NCAR), Laure Zanna (NYU), Sara Shamekh (NYU), Tian Zheng (Columbia University), Robert King (Stanford), Akshay Subramaniam (NVIDIA), and Stephan Hoyer (Google) for their insightful discussions during the 2024 LEAP-STC NSF site visit, the 2024 LEAP-STC annual meeting, and AGU24 Annual Meeting. Finally, we would like to express our gratitude to the anonymous reviewers at JAMES for their meticulous, constructive, and thoughtful feedback.

\bibliography{main}
\clearpage

\title{Supporting Information for ``Navigating the Noise: Bringing Clarity to ML Parameterization Design with $\mathcal{O}$(100) Ensembles"}

\authors{Jerry Lin$^1$, Sungduk Yu$^2$, Liran Peng$^1$, Tom Beucler$^{3,4}$, Eliot Wong-Toi$^5$, Zeyuan Hu$^{6,7}$, Pierre Gentine $^{8}$, Margarita Geleta $^{9,10}$, Mike Pritchard$^{1,7}$}

\affiliation{1}{Department of Earth System Sciences, University of California at Irvine, Irvine, CA, USA}
\affiliation{2}{Intel Corporation}
\affiliation{3}{Faculty of Geosciences and Environment, University of Lausanne, Lausanne, Switzerland}
\affiliation{4}{Expertise Center for Climate Extremes, University of Lausanne, Lausanne, Switzerland}
\affiliation{5}{Department of Statistics, University of California at Irvine, Irvine, CA, USA}
\affiliation{6}{Department of Earth and Planetary Sciences, Harvard University}
\affiliation{7}{NVIDIA Research}
\affiliation{8}{LEAP Science and Technology Center, Columbia University}
\affiliation{9}{Department of Electrical Engineering and Computer Science, University of California at Berkeley, Berkeley, CA, USA}
\affiliation{10}{Department of Biomedical Data Science, Stanford University School of Medicine, Palo Alto, CA, USA}

%
%

%


%
%
\renewcommand{\thefigure}{S\arabic{figure}}
\renewcommand{\thetable}{S\arabic{table}}
\setcounter{figure}{0}
\setcounter{table}{0}

\noindent\textbf{Contents of this file}

\begin{enumerate}

\item Text S1 to S3

\item Tables S1 to S2

\item Figures S1 to S35

\end{enumerate}

\noindent\textbf{Introduction}

Text S1 provides additional detail regarding preprocessing and imperfections in SP-CAM3 simulation data, and Text S2 explains how offline and online error were calculated. Text S3 provides more details on sample size estimation. Table S1 is analogous to Table 3 in the main manuscript, except we report statistics for offline moistening and online moisture instead of offline heating and online temperature. Table S2 shows offline heating and moistening error using a multiclimate, out-of-distribution test set along with the corresponding Spearman correlations to ``in-distribution" online error. In this context, in-distribution refers to the fact that the prescribed SSTs are the same in offline training and online testing.  Figure S1 shows the Benjamini-Hochberg correction being applied to all hypothesis tests conducted in this study. Figure S2 shows the percent change in in-distribution offline test error with a doubling and quadrupling in size for the offline test set. Figure S3 is analogous to Figure 2 in the main text, except the offline error curves are for the multiclimate, out-of-distribution test set. Figures S4 to S12 show online error growth across configurations, colorized by offline error. Figure S13 (S14) shows median (mean) online error growth for each configuration. Figure S15 shows histograms of number of months completed without crashing for hybrid simulations across each configuration. Figures S16 to S24 show offline error vs. online error for all configurations. Figure S25 shows offline vs. online error boxplots for all configurations. Figures S26, S27, S28, S29, S30, and S31 show the effect of dropout, leak (in Leaky ReLu), learning rate, parameter count, choice of optimizer, and batch normalization on offline and online error across all NNs that did not crash after marginalizing out choice of configuration. Figure S32 shows the effects of the aforementioned model properties on stability after marginalizing out choice of configuration. Figure S33 shows mean online temperature and moisture error growth for hybrid simulations with and without batch normalization. Figures S34 and S35 show Laplacian distributions being fit to histograms of the heating and moistening tendencies in the training data, respectively.

\noindent\textbf{Text S1.}

Without further preprocessing, the formula for number of samples for the training data after subsampling longitudinally would be  $N_{samples} = N_{days} \times 48 \times 64 \times 128 \times \frac{1}{8}$ where 48 corresponds to timesteps per day, 64 corresponds to number of latitudes, 128 corresponds to the number of longitudes, and 1/8 corresponds to longitudinal subsampling. The formula for validation data is similar; however 19 longitudes are subsampled as 128 is not divisible by 7. Additionally, the final timesteps of every month and the simulation itself are removed because of strange behavior at the final timestep of SP-CAM3 simulations and the modular structure of the preprocessing code.

This yields $365 \times 48 \times 64 \times 128 \times (1/8) - 13 \times 64 \times 128 \times (1/8) = 17,927,168$ samples for the training data and $61 \times 48 \times 64 \times 19 - 3 * 64 * 19 = 3,556,800$ samples for the validation data. Since the offline test data is not subsampled, the number of samples becomes $7 \times 48 \times 64 \times 128 = 2,752,512$ samples for offline test data. For the multiclimate configuration, months 2 through 5 were similarly subsampled across -4K, +0K, and +4K climates. To control for the data volume, only the first 5,975,723 samples were kept for each climate. Although this yields one additional sample in the training data compared to the other configurations, we do not believe the addition of a single example significantly impacts our results. 

\noindent\textbf{Text S2.}

Both offline and online error are mass-weighted using area and pressure (from the SPCAM3 reference simulation). However, offline RMSE is calculated across all samples while online RMSE error is first calculated on a simulation-month basis and, for the main manuscript, averaged across all months for runs that do not crash.

Offline heating and moistening RMSE are both defined as:

\[
\sqrt{\sum_{i} W_i \circ (\hat{Y}_i - Y_i)^2}
\]

where:
\begin{itemize}
    \item \( \circ \) denotes element-wise multiplication.
    \item \( (\cdot)^2 \) indicates squaring each element individually.
    \item \( \hat{Y} \in \mathbb{R}^{336 \times 30 \times 64 \times 128} \): the NN prediction tensor.
    \item \( Y \in \mathbb{R}^{336 \times 30 \times 64 \times 128} \): the ground truth tensor.
    \item \( W \in \mathbb{R}^{336 \times 30 \times 64 \times 128} \): the weight tensor, whose values sum to one.
    \item \( i \): the flattened index over all dimensions of the tensors: time (336 timesteps), vertical levels (30), latitude (64), and longitude (128).
\end{itemize}

Online temperature and moisture RMSE are both defined as:

$$\sqrt{\sum\limits_{i} W_i \circ (Error^*_i)^2}$$
where:
\[
\text{Error}[m] = 
\frac{
\sum_{\text{lon}=1}^{8} \sum_{t=1}^{T_m} \hat{y}_{m, t, \text{lon}} 
}{8 \cdot T_m}
- 
\frac{
\sum_{\text{lon}=1}^{8} \sum_{t=1}^{T_m} y_{m, t, \text{lon}} 
}{8 \cdot T_m}, \quad \text{for } m = 1, \ldots, 12
\]
and:
\begin{itemize}
\item $\hat{y}_{m,t,lon}$ refers to coupled NN simulation output for a given month, day of month, and longitude
\item $y_{m,t,lon}$ refers to reference SPCAM 3 simulation output for a given month, day of month, and longitude
\item $T_m$ denotes number of days in month $m$
\item \( \circ \) denotes element-wise multiplication.
\item \( (\cdot)^2 \) indicates squaring each element individually.
\item Error[m] refers to the unweighted online error for a given month $m$, averaged over eight longitudes and $T_m$ days.
\item $Error^* \in \mathbb{R}^{N_M \times 30 \times 64}$: the error tensor with $N_M$ denoting number of months fully integrated without crashing. It is formed by concatenating monthly errors (i.e. Error[m] for a given month) along the time dimension.
\item $W \in \mathbb{R}^{N_M \times 30 \times 64}$: the mass-weight tensor that, when summed along the vertical level and latitude dimensions, results in a vector of ones $\in \mathbb{R}^{N_M}$. Again, $N_m$ denotes the number of months fully integrated without crashing.
\item \( i \): the flattened index over the vertical level and latitude dimensions of the tensors
\item the final result is a vector $\in \mathbb{R}^{12}$
\end{itemize}

Only eight longitudes are used for online error because every 16th longitude is automatically subsampled in the semi-automated pipeline to mitigate storage consumption. For online simulations that crash, the formula is the same except the time dimension for both the weight and error tensors is limited to the number of months completed without crashing.

\noindent\textbf{Text S3.}
In line with work from \citeA{Noether1987-po} and \citeA{Hamilton1991-xa} for power calculations for Mann Whitney U-Tests, we use the following formula to estimate necessary sample sizes:

$$n^* = \left\lceil \left\lceil \frac{\left(\Phi^{-1}(\kappa_0) + \Phi^{-1}(1 - \alpha)\right)^2}{6(P-.5)^2} \right\rceil * \frac{1}{\gamma} \right\rceil$$

where $n^*$ is our sample size estimate, $\kappa_0 = .8$ is our desired power, $\alpha = .05$ is our significance level, $\gamma = 141/330$ is our assumed survival rate (i.e., the surival rate of the standard configuration), and $P$ is $P(Y > X)$, where $X$ and $Y$ are realizations from the populations that our configurations are each sampling from. $\Phi^{-1}$ is the inverse of the cumulative distribution function of a normal distribution, and the ceiling notation indicates rounding up to the next integer. In the context of the sample size estimate that uses the difference in online error when omitting ozone, we get a U-statistic of 8,363. This means $P \approx 42.1\%$ and $n^* \approx 384$. 

\begin{table}[h]
\small
\centering
\begin{tabular}{l c c c c c c c}
\hline
\textbf{Configuration} & \textbf{Survival} & \textbf{Online Error (g/kg)} & \textbf{Offline Error (g/kg/day)} & \textbf{Spearman's $\rho$}\\
\hline
standard          & 42.7\% & .423 g/kg & 1.76 g/kg/day & \textbf{.811} \\

specific humidity & \textcolor{red}{\textbf{-26.1\%}} & +.032 g/kg & \textcolor{red}{\textbf{+.00789 g/kg/day}} & \textbf{.806} \\

no memory         & \textcolor{red}{\textbf{-32.7\%}} & \textcolor{red}{\textbf{+.246 g/kg}} & \textcolor{red}{\textbf{+.256 g/kg/day}} & .306 \\

no wind           & +1.82\% & -.004 g/kg & +.00412 g/kg/day & \textbf{.716} \\

no ozone          & \textcolor{teal}{\textbf{+13.3\%}} & +.0354 g/kg & +.00252 g/kg/day & \textbf{.847} \\

no zenith angle   & \textcolor{teal}{\textbf{+11.2\%}} & -.0357 g/kg & +.00134 g/kg/day & \textbf{.798} \\

MAE               & \textcolor{teal}{\textbf{+32.4\%}} & \textcolor{red}{\textbf{+.335 g/kg}} & \textcolor{teal}{\textbf{-.00952 g/kg/day}} & \textbf{.790} \\

no dropout        & \textcolor{red}{\textbf{-14.2\%}} & \textcolor{teal}{\textbf{-.0850 g/kg}} & \textcolor{teal}{\textbf{-.0545 g/kg/day}} & .194 \\

multiclimate      & \textcolor{teal}{\textbf{+27.3\%}} & \textcolor{teal}{\textbf{-.0535 g/kg}} & -.00360  g/kg/day & \textbf{.773} \\
\hline

\end{tabular}
\caption{This table is analogous to Table 3 in the main manuscript except we report statistics for offline moistening error and online moisture error instead of offline heating error and online temperature error. \label{tab:summarytable_Q}}
\end{table}

\begin{table}[h]
\small
\centering
\begin{tabular}{l c c c c c c c}
\hline
\textbf{Configuration} & \textbf{Offline Error\textsuperscript{*} (T)} & \textbf{$\rho^*$ (T)} & \textbf{Offline Error\textsuperscript{*} (Q)} & \textbf{$\rho^*$ (Q)}\\
\hline
standard          & 2.04 K/day & \textbf{.698} & 1.85 g/kg/day & \textbf{.734} \\

specific humidity & \textcolor{red}{\textbf{+1.54 K/day}} & \textbf{.331} & \textcolor{red}{\textbf{+.393 g/kg/day}} & \textbf{.304} \\

no memory         & \textcolor{red}{\textbf{+.872 K/day}} & .250 & \textcolor{red}{\textbf{+.395 g/kg/day}} & .0842 \\

no wind           & $+5.43 \times 10^{-5}$ K/day & \textbf{.624} & +.00445 g/kg/day & \textbf{.585} \\

no ozone          & +.00259 K/day & \textbf{.698} & -.00534 g/kg/day & \textbf{.846} \\

no zenith angle   & +.00561 K/day & \textbf{.659} & +.00267 g/kg/day & \textbf{.708} \\

MAE               & \textcolor{teal}{\textbf{-0.0575 K/day}} & \textbf{.688} & \textcolor{teal}{\textbf{-.0258 g/kg/day}} & \textbf{.783} \\

no dropout        & \textcolor{teal}{\textbf{-0.0259 K/day}} & .186 & \textcolor{red}{\textbf{+.0777 g/kg/day}} & .0752 \\

multiclimate      & \textcolor{teal}{\textbf{-0.0842 K/day}} & \textbf{.860} & \textcolor{teal}{\textbf{-.0405 g/kg/day}} & \textbf{.774} \\
\hline

\end{tabular}
\caption{This table shows offline error for heating in K/day, Spearman correlation between offline heating and online temperature RMSE, offline error for moistening in g/kg/day, and Spearman correlation between offline moistening and online moisture RMSE using a multiclimate offline test set. As expected, the multiclimate configuration performs the best on this test set. The significantly worse performance for the specific humidity configuration here also confirms the ``climate-invariant'' benefits of a relative humidity transformation for moisture input first shown in \citeA{Beucler2024-vb}.}
\label{tab:summarytable_multiclimate}
\end{table}

\begin{figure}[!htbp]
 \centering
 \includegraphics[width=\textwidth]{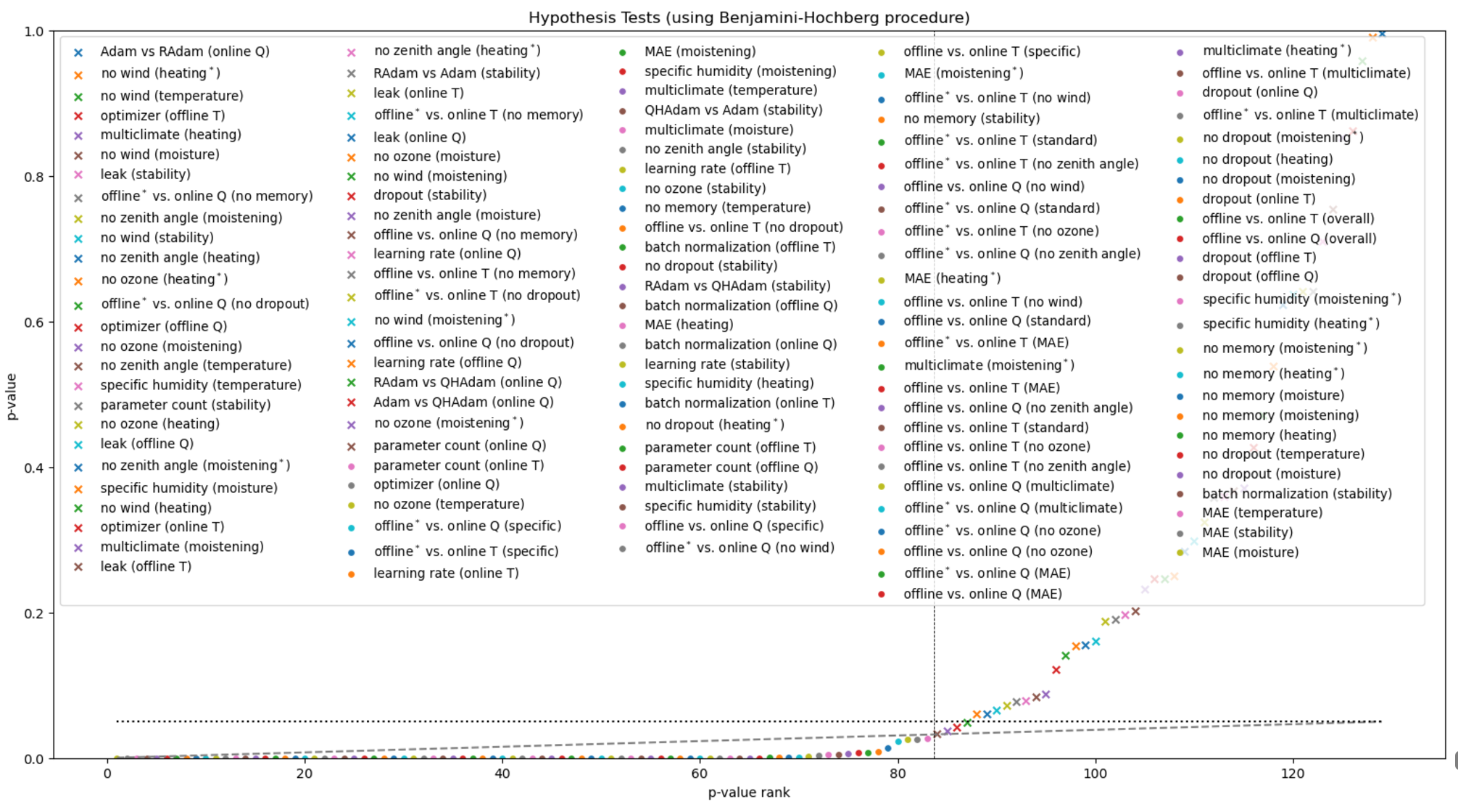}

\caption{This figure illustrates the Benjamini-Hochberg correction being applied to ranked unadjusted p-values. Dots to the left of the dashed line correspond to statistically significant p-values. The p-value associated with the permutation test comparing online temperature error from the standard and no ozone configurations is the largest statistically significant p-value. Heating and moistening correspond to offline error while temperature and moisture correspond to online error. Labels with an asterisk indicate inference on a multiclimate offline test set.}
 \label{fig:testcorrection}
\end{figure}

\begin{figure}[!htbp]
 \centering
 \includegraphics[width=\textwidth]{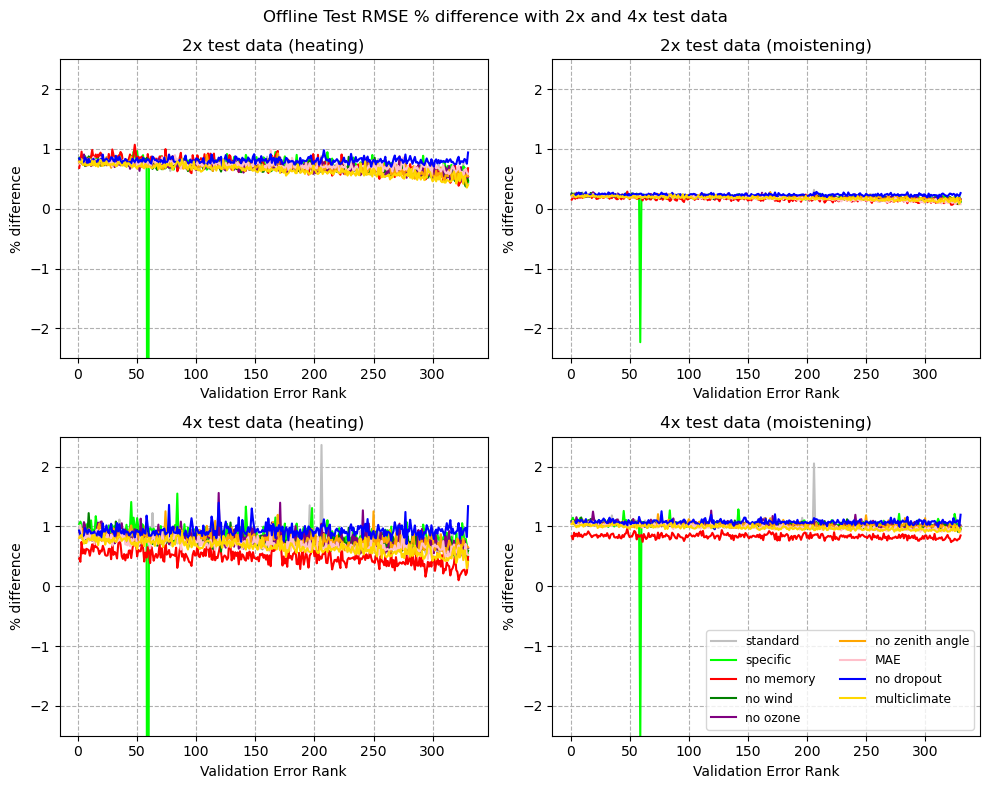}

\caption{This shows the percent change in offline test heating and moistening RMSE when doubling and quadrupling the test data.}
 \label{fig:testdata2x4x}
\end{figure}

\begin{figure}[!htbp]
 \centering
 \includegraphics[width=\textwidth]{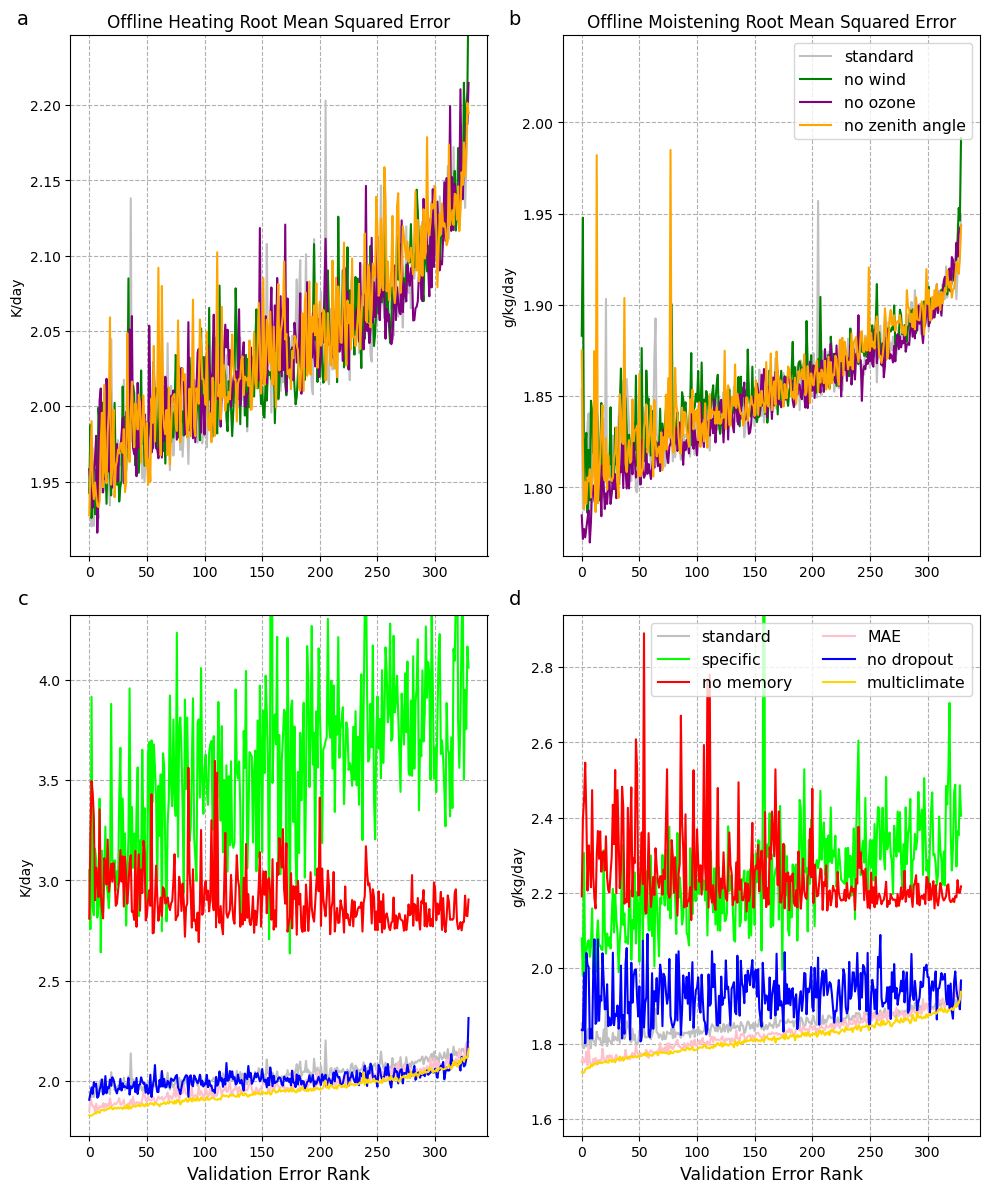}

\caption{This figure is analogous to Figure 2 in the main text, showing the offline RMSE for all 2,970 models from nine configurations on a multiclimate test set. S2c and S2d show configurations with statistically distinct average RMSE from the standard configuration.}
 \label{fig:Offline_Error_Multiclimate}
\end{figure}

\begin{figure}[!htbp]
 \centering
 \includegraphics[width=\textwidth]{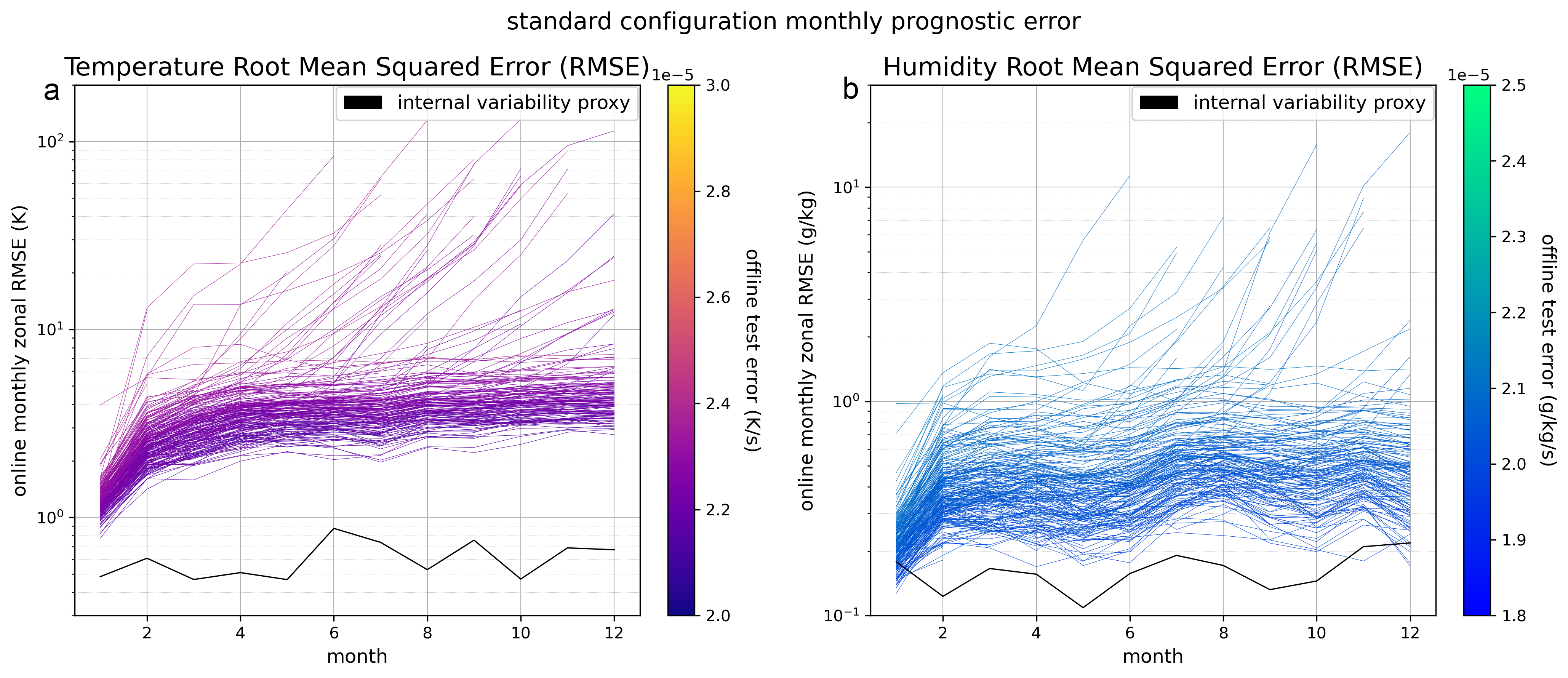}

\caption{This figure shows online error over time across all 330s NNs trained in the standard configuration. An internal variability proxy that is calculated via the error of a 31-day lagged version of the reference simulation is included for comparison.}
 \label{fig:online_diffs_standard}
\end{figure}

\begin{figure}[!htbp]
 \centering
 \includegraphics[width=\textwidth]{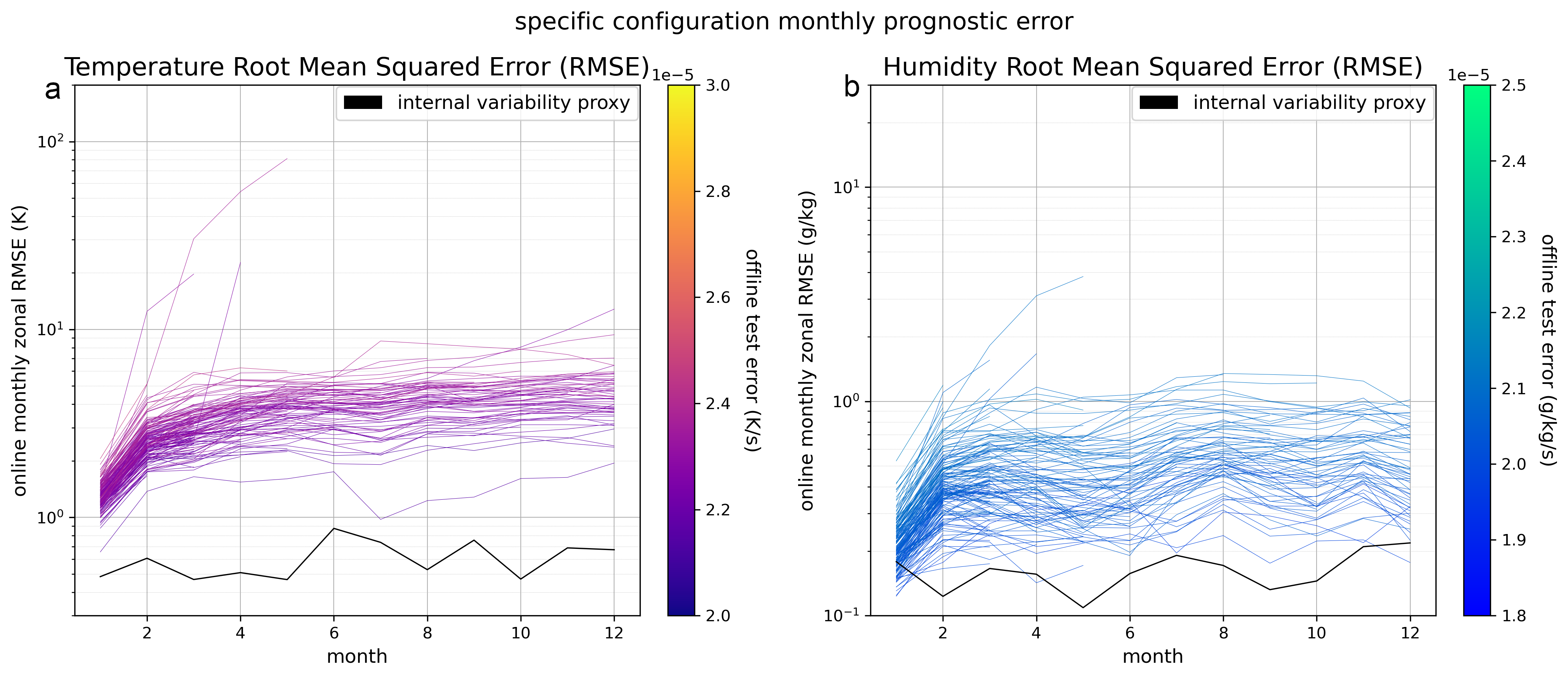}

\caption{This figure shows online error over time across all 330s NNs trained in the specific configuration. An internal variability proxy that is calculated via the error of a 31-day lagged version of the reference simulation is included for comparison.}
 \label{fig:online_diffs_specific}
\end{figure}

\begin{figure}[!htbp]
 \centering
 \includegraphics[width=\textwidth]{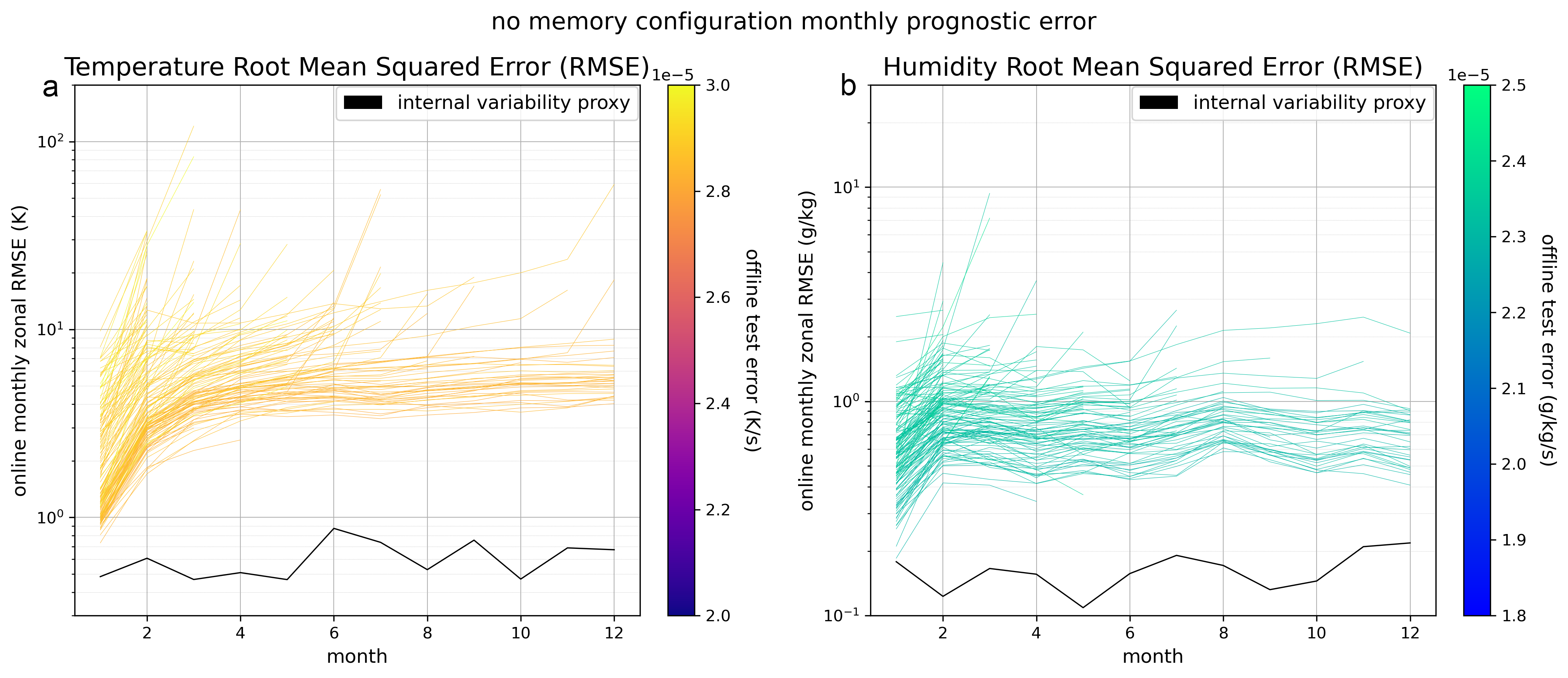}

\caption{This figure shows online error over time across all 330s NNs trained in the no memory configuration. An internal variability proxy that is calculated via the error of a 31-day lagged version of the reference simulation is included for comparison.}
 \label{fig:online_diffs_nomemory}
\end{figure}

\begin{figure}[!htbp]
 \centering
 \includegraphics[width=\textwidth]{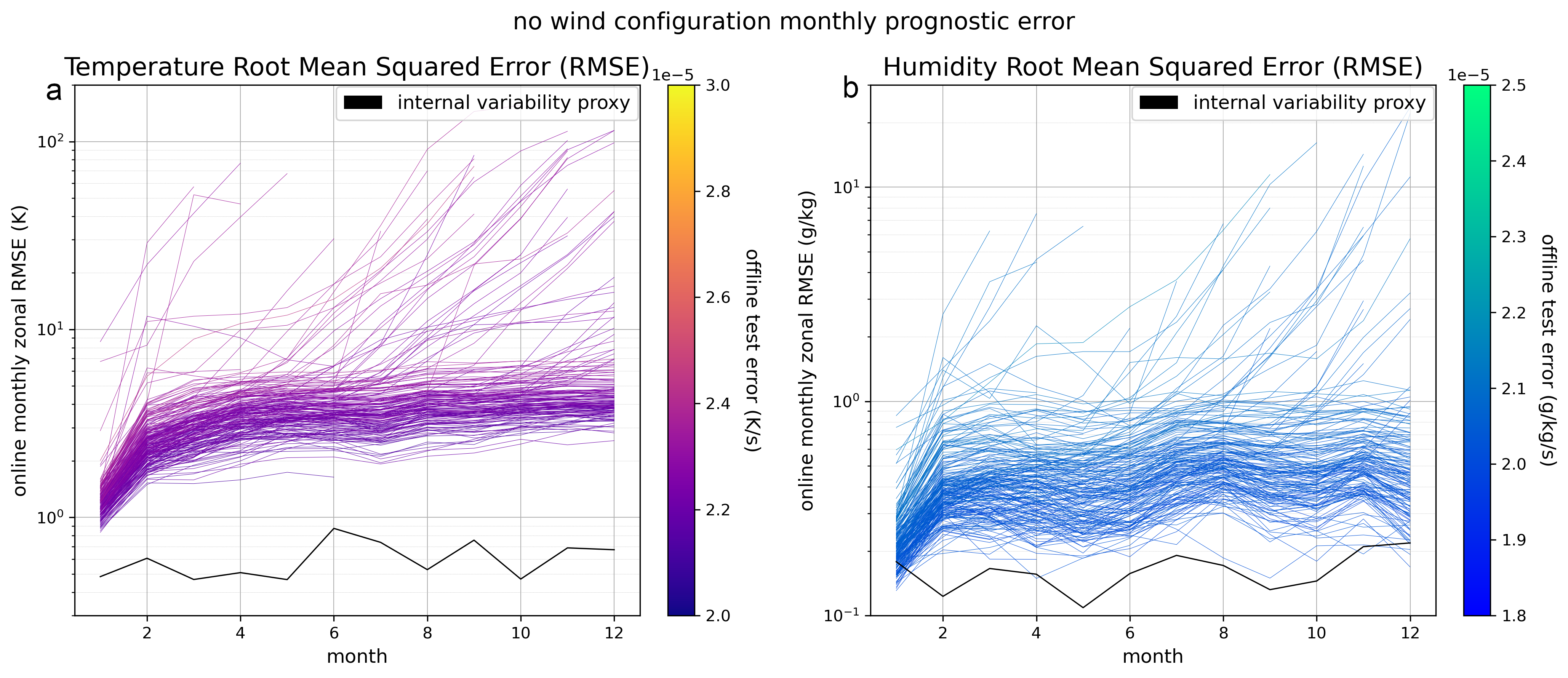}

\caption{This figure shows online error over time across all 330s NNs trained in the no wind configuration. An internal variability proxy that is calculated via the error of a 31-day lagged version of the reference simulation is included for comparison.}
 \label{fig:online_diffs_nowind}
\end{figure}

\begin{figure}[!htbp]
 \centering
 \includegraphics[width=\textwidth]{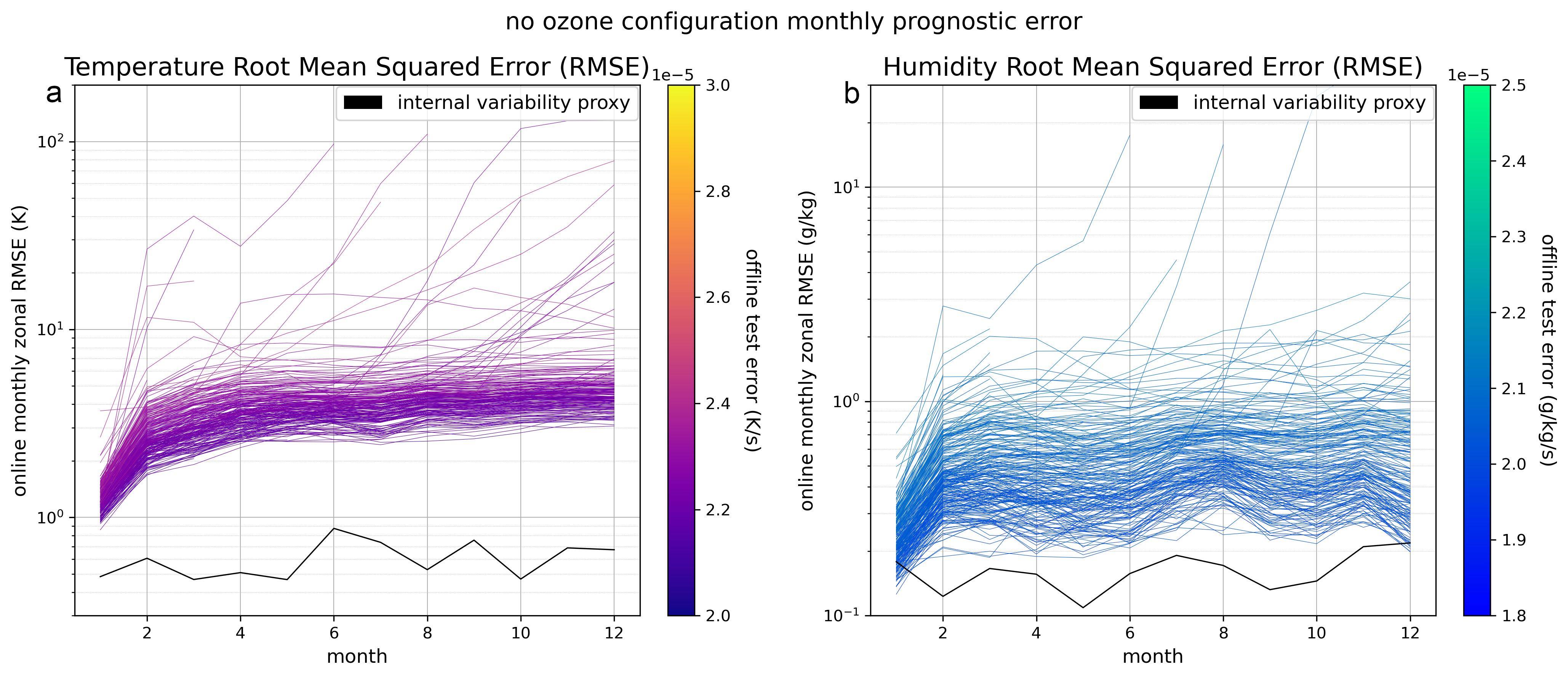}

\caption{This figure shows online error over time across all 330s NNs trained in the no ozone configuration. An internal variability proxy that is calculated via the error of a 31-day lagged version of the reference simulation is included for comparison.}
 \label{fig:online_diffs_noozone}
\end{figure}

\begin{figure}[!htbp]
 \centering
 \includegraphics[width=\textwidth]{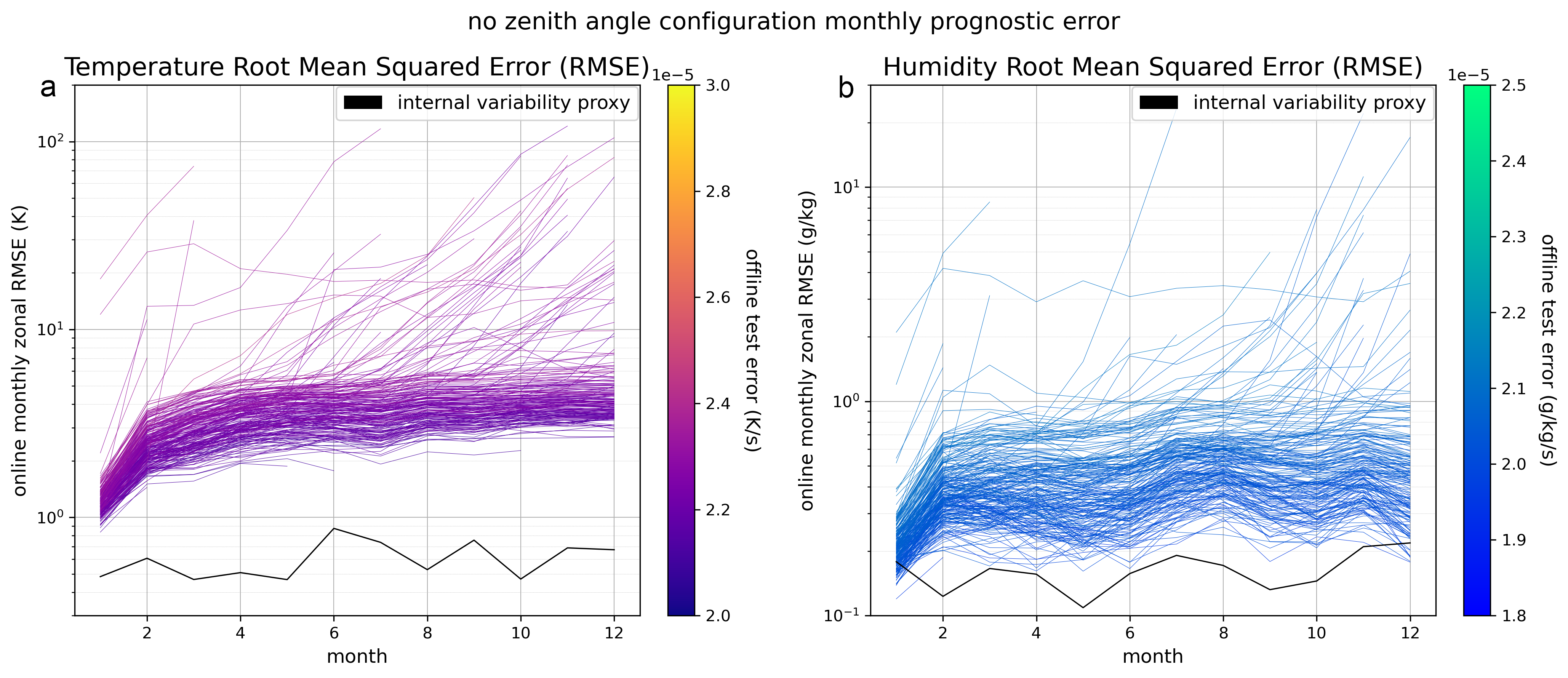}

\caption{This figure shows online error over time across all 330s NNs trained in the no zenith angle configuration. An internal variability proxy that is calculated via the error of a 31-day lagged version of the reference simulation is included for comparison.}
 \label{fig:online_diffs_nocoszrs}
\end{figure}

\begin{figure}[!htbp]
 \centering
 \includegraphics[width=\textwidth]{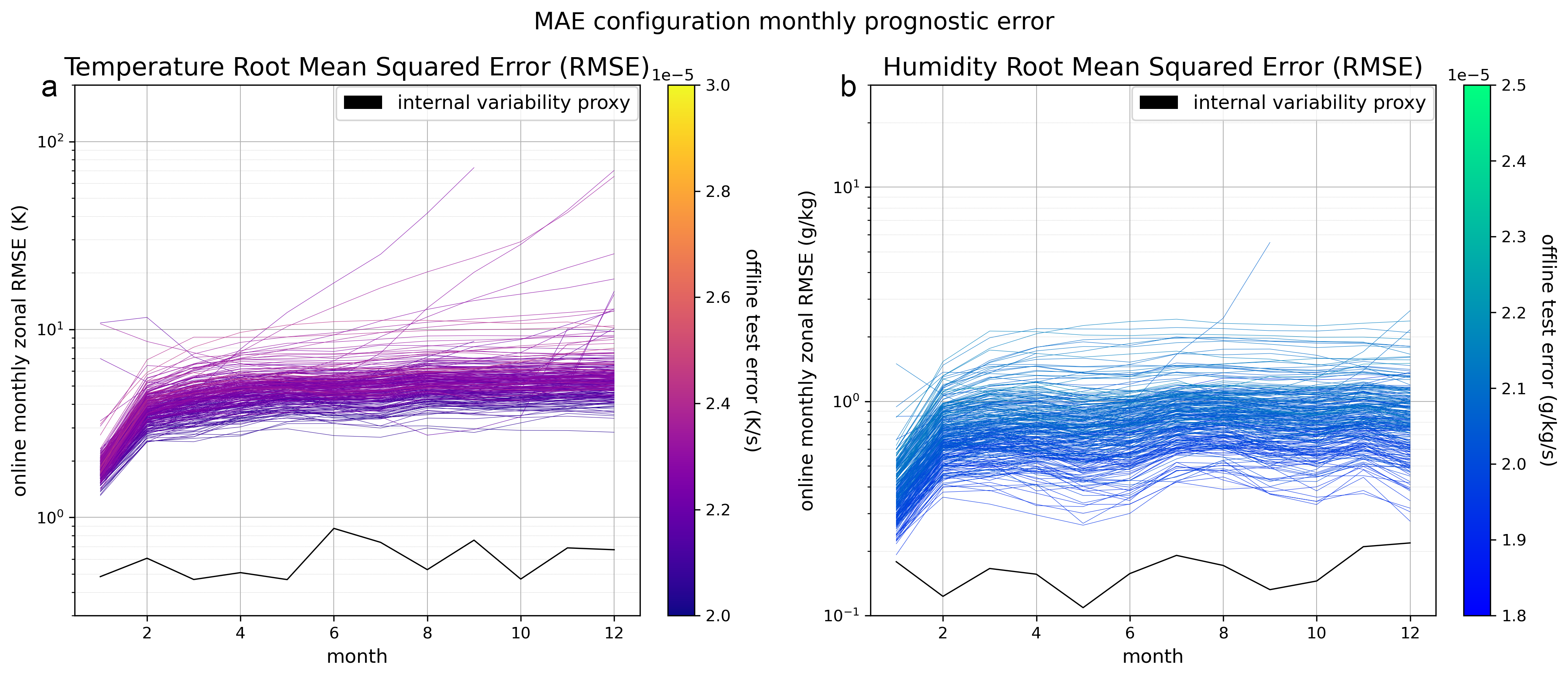}

\caption{This figure shows online error over time across all 330s NNs trained in the MAE configuration. An internal variability proxy that is calculated via the error of a 31-day lagged version of the reference simulation is included for comparison.}
 \label{fig:online_diffs_mae}
\end{figure}

\begin{figure}[!htbp]
 \centering
 \includegraphics[width=\textwidth]{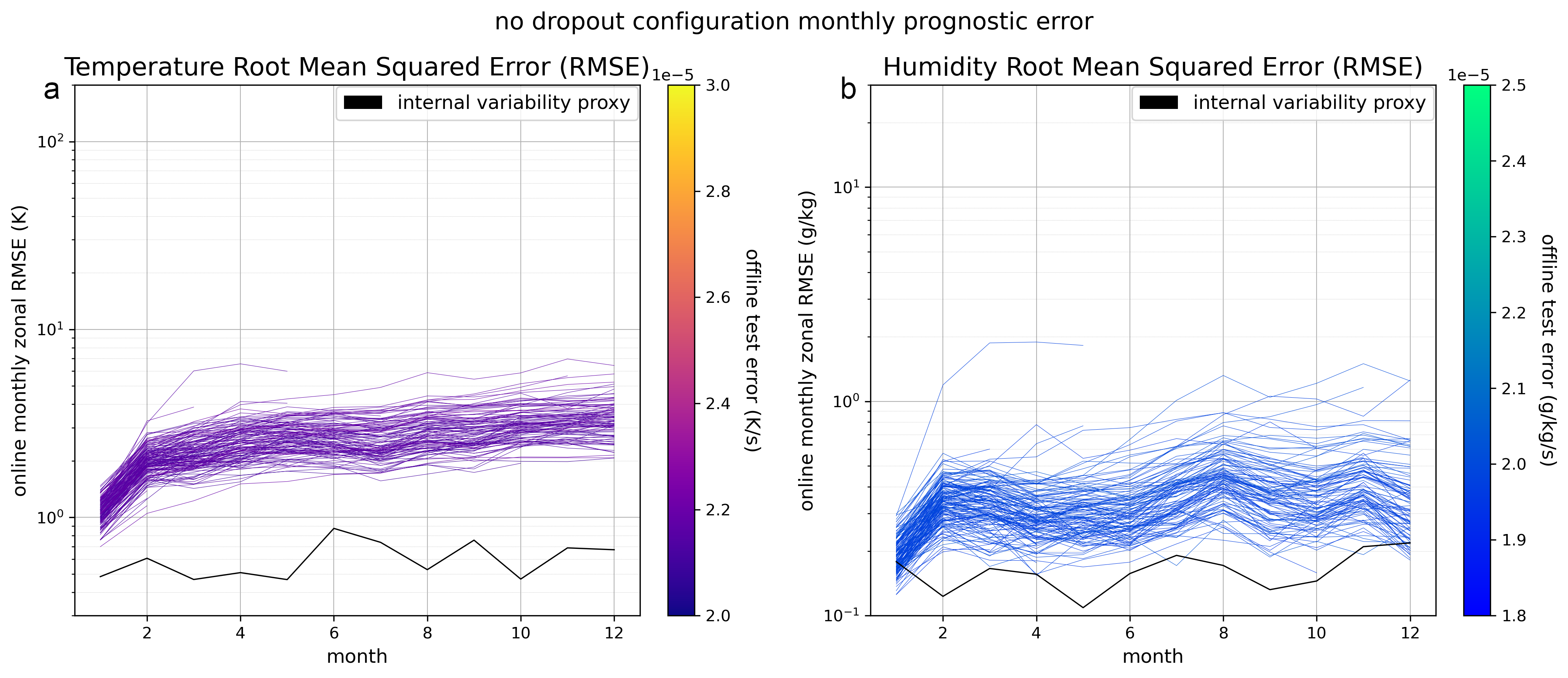}

\caption{This figure shows online error over time across all 330s NNs trained in the no dropout configuration. An internal variability proxy that is calculated via the error of a 31-day lagged version of the reference simulation is included for comparison.}
 \label{fig:online_diffs_nodropout}
\end{figure}

\begin{figure}[!htbp]
 \centering
 \includegraphics[width=\textwidth]{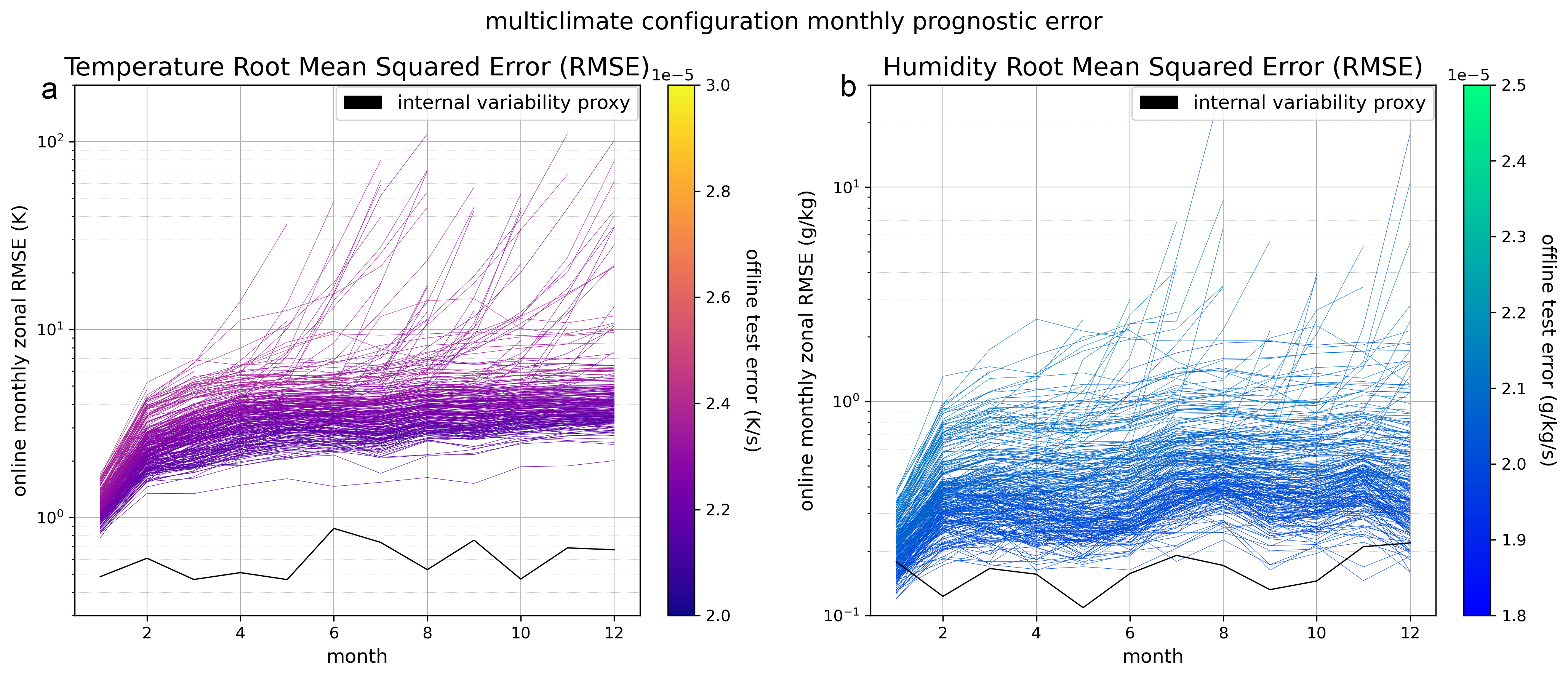}

\caption{This figure shows online error over time across all 330s NNs trained in the multiclimate configuration. An internal variability proxy that is calculated via the error of a 31-day lagged version of the reference simulation is included for comparison.}
 \label{fig:online_diffs_multiclimate}
\end{figure}

\begin{figure}[!htbp]
 \centering
 \includegraphics[width=\textwidth]{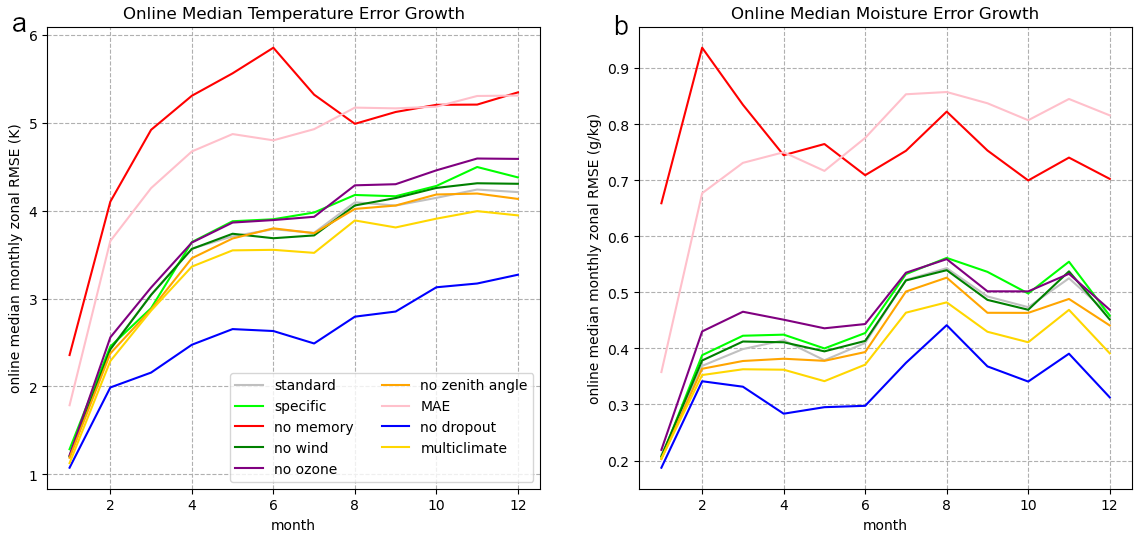}

\caption{This figure shows median online error for both temperature and moisture over time for each configuration. Ensemble-median online error is only recorded for simulations that do not crash, and they are calculated across the entire duration of the simulations.}
 \label{fig:median_error_growth}
\end{figure}

\begin{figure}[!htbp]
 \centering
 \includegraphics[width=\textwidth]{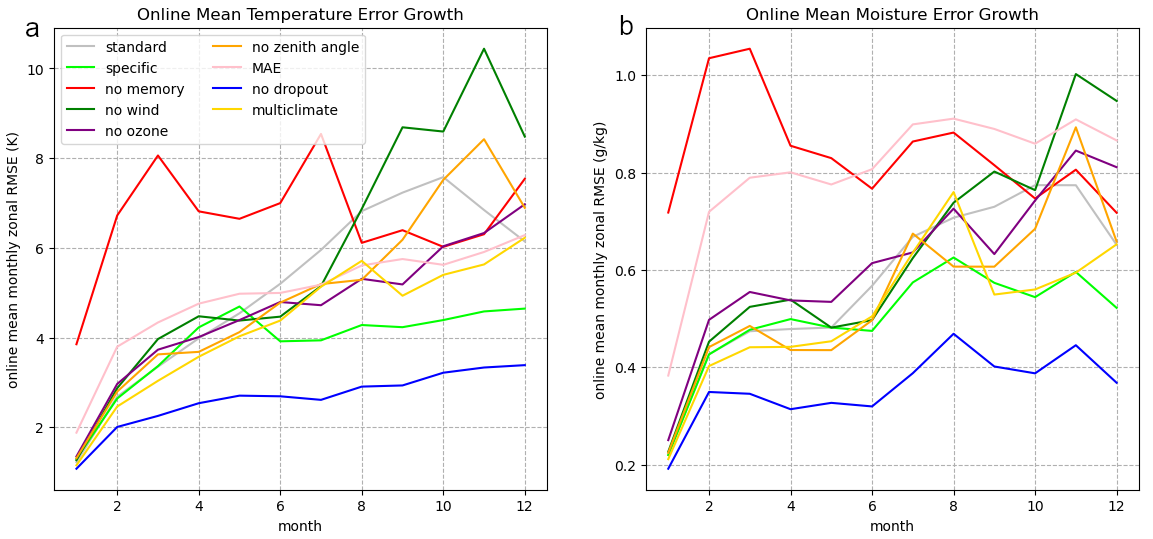}

\caption{This figure shows mean online error for both temperature and moisture over time for each configuration. Ensemble-mean online error is only recorded for simulations that do not crash, and they are calculated across the entire duration of the simulations.}
 \label{fig:mean_error_growth}
\end{figure}

\begin{figure}[!htbp]
 \centering
 \includegraphics[width=\textwidth]{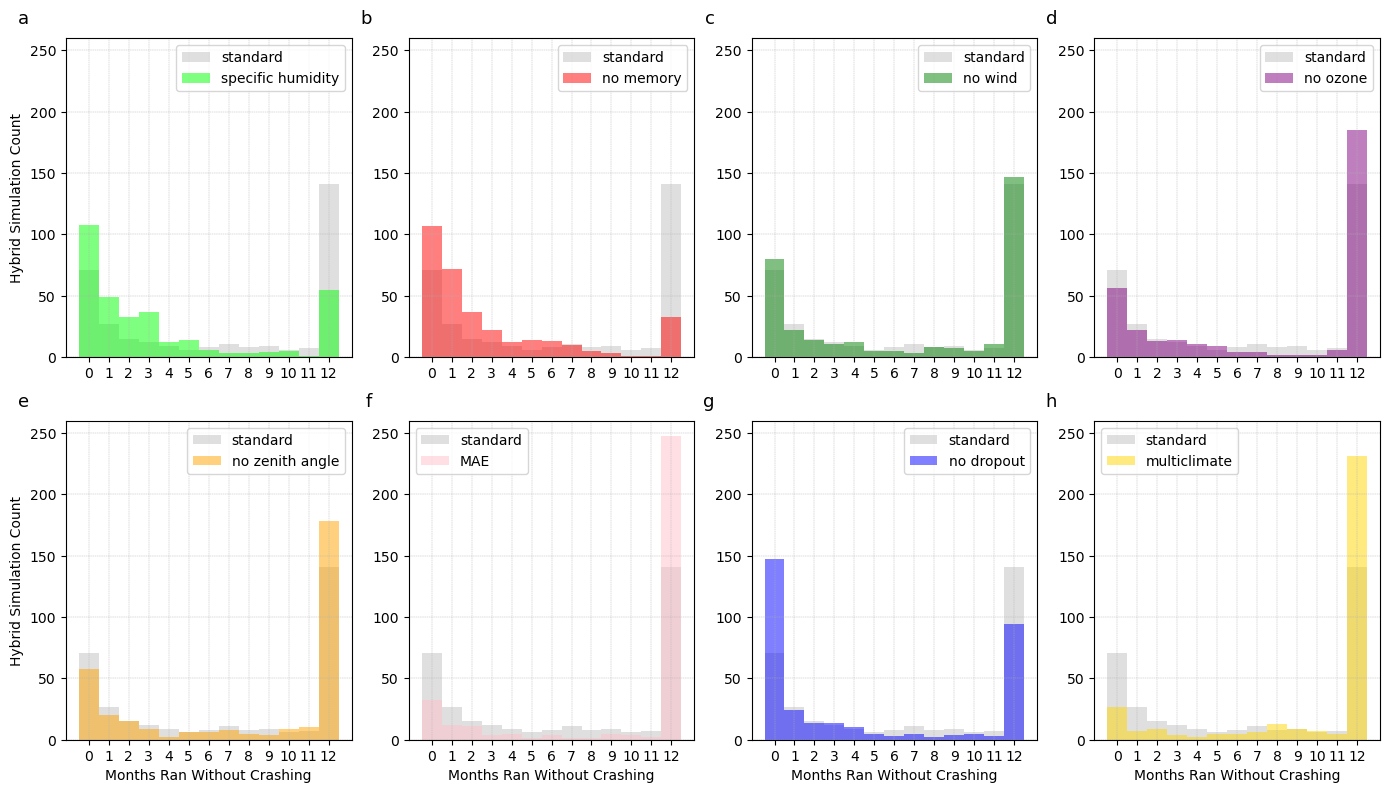}

\caption{This figure shows the number of simulation months integrated for each configuration compared to the standard configuration.}
 \label{fig:months_ran}
\end{figure}

\begin{figure}[!htbp]
 \centering
 \includegraphics[width=\textwidth]{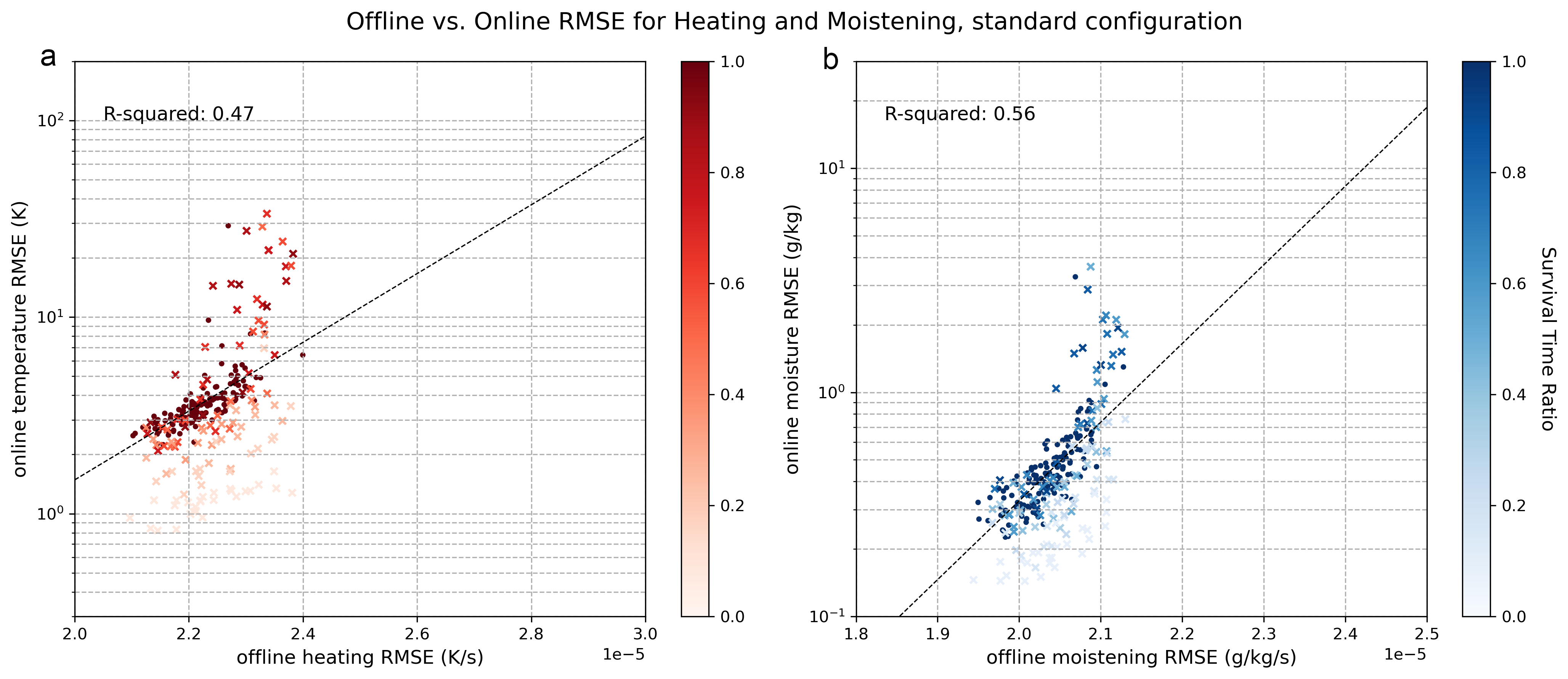}

\caption{This figure shows offline error vs. online error across all 330 models in the standard configuration. Models that crash mid-simulation are denoted with an `x'. A linear regression is fit using offline error to predict log-transformed online error for NNs that do not crash, and the corresponding Pearson's R-squared is shown in the top left-hand corner.}
 \label{fig:offlinevonline_standard}
\end{figure}

\begin{figure}[!htbp]
 \centering
 \includegraphics[width=\textwidth]{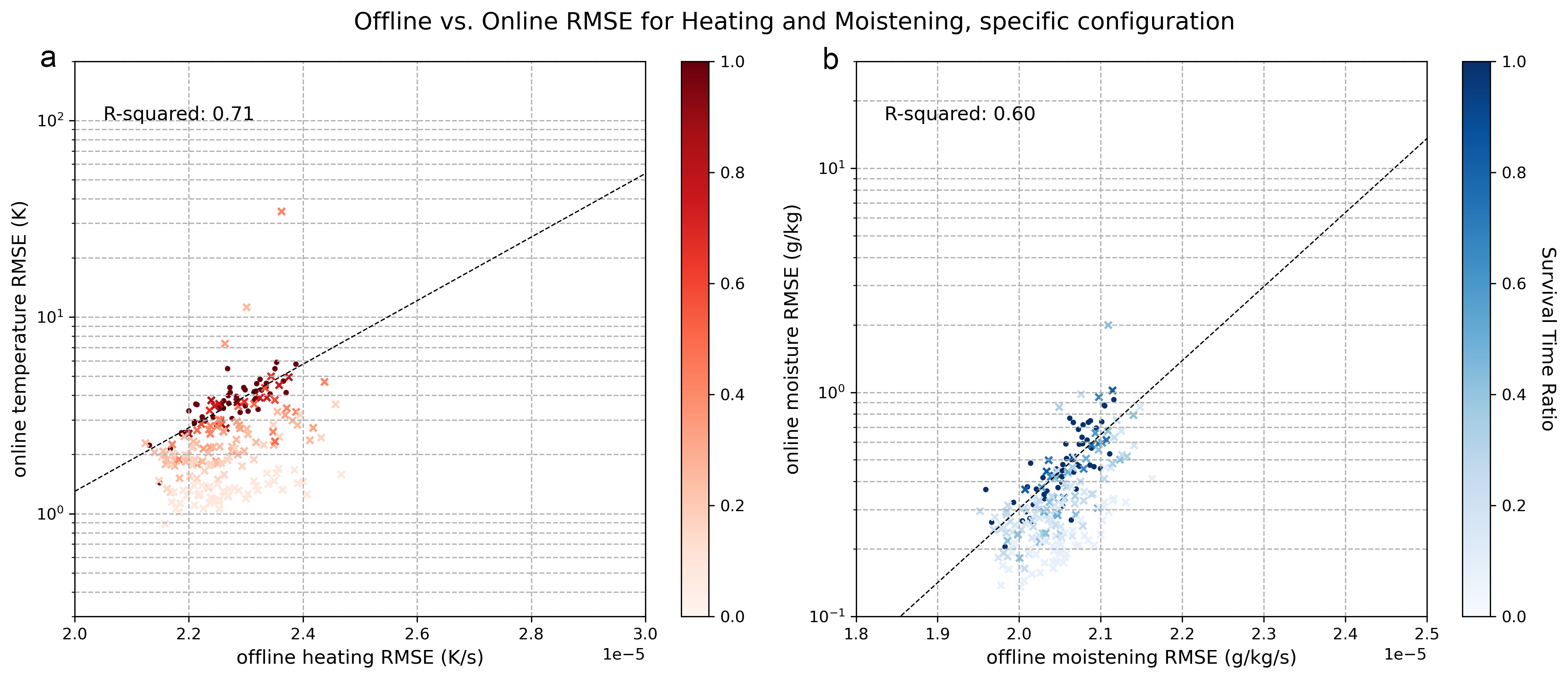}

\caption{This figure shows offline error vs. online error across all 330 models in the specific configuration. Models that crash mid-simulation are denoted with an `x'. A linear regression is fit using offline error to predict log-transformed online error for NNs that do not crash, and the corresponding Pearson's R-squared is shown in the top left-hand corner.}
 \label{fig:offlinevonline_specific}
\end{figure}

\begin{figure}[!htbp]
 \centering
 \includegraphics[width=\textwidth]{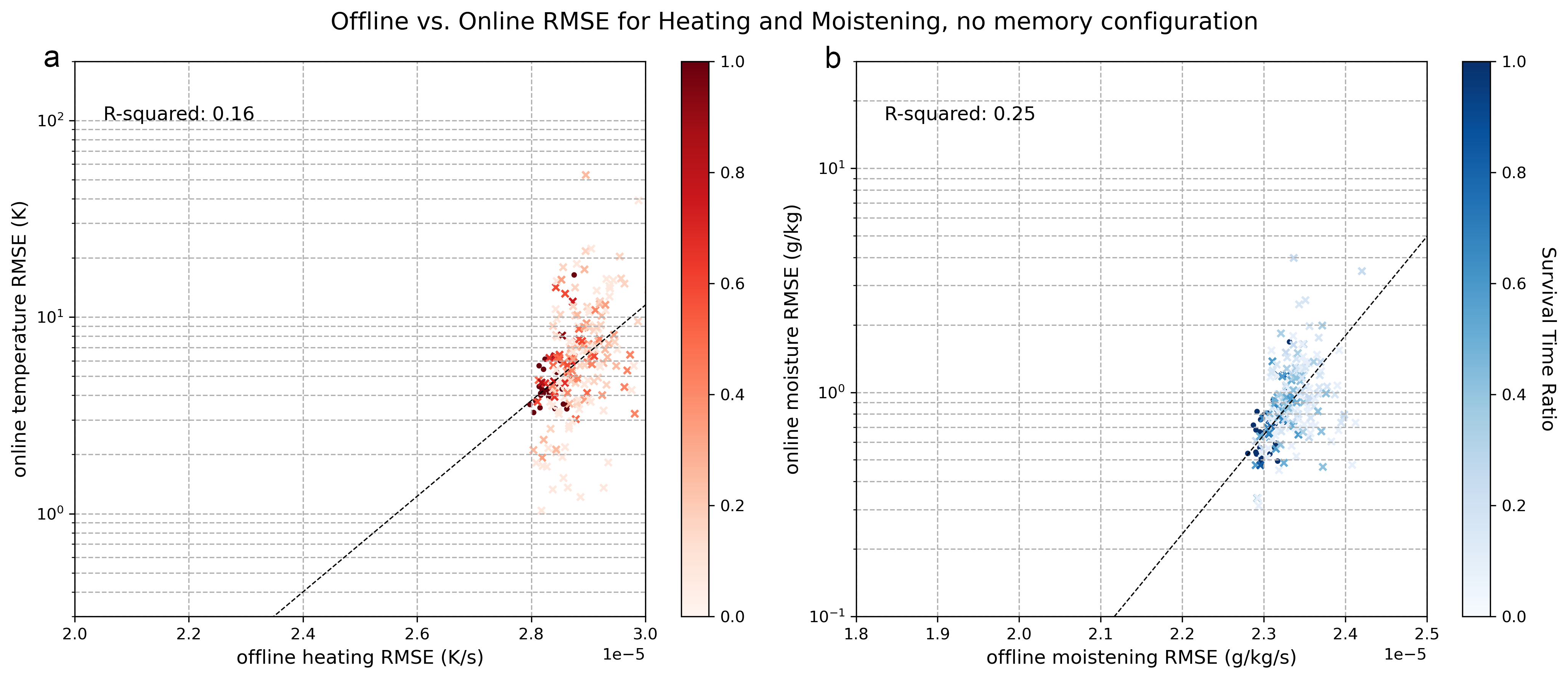}

\caption{This figure shows offline error vs. online error across all 330 models in the no memory configuration. Models that crash mid-simulation are denoted with an `x'. A linear regression is fit using offline error to predict log-transformed online error for NNs that do not crash, and the corresponding Pearson's R-squared is shown in the top left-hand corner.}
 \label{fig:offlinevonline_nomemory}
\end{figure}

\begin{figure}[!htbp]
 \centering
 \includegraphics[width=\textwidth]{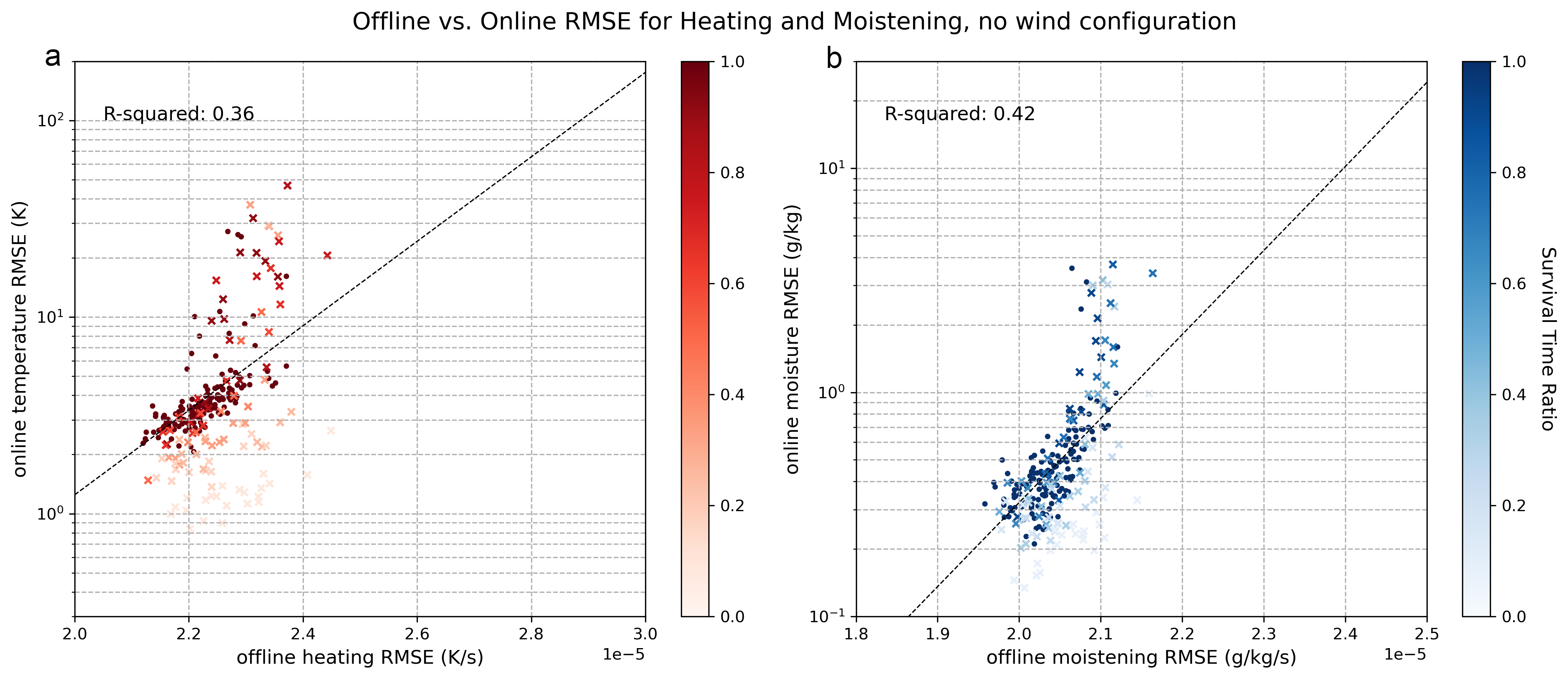}

\caption{This figure shows offline error vs. online error across all 330 models in the no wind configuration. Models that crash mid-simulation are denoted with an `x'. A linear regression is fit using offline error to predict log-transformed online error for NNs that do not crash, and the corresponding Pearson's R-squared is shown in the top left-hand corner.}
 \label{fig:offlinevonline_nowind}
\end{figure}

\begin{figure}[!htbp]
 \centering
 \includegraphics[width=\textwidth]{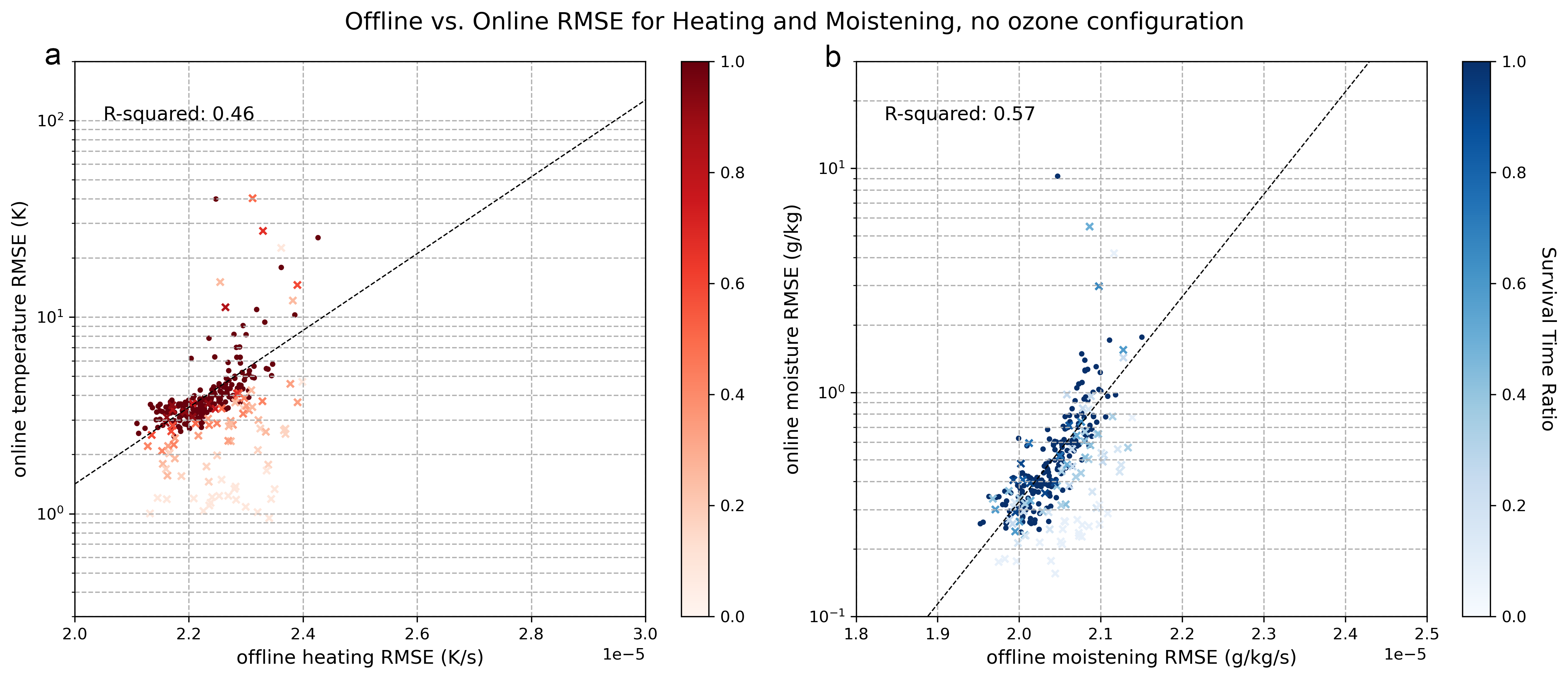}

\caption{This figure shows offline error vs. online error across all 330 models in the no ozone configuration. Models that crash mid-simulation are denoted with an `x'. A linear regression is fit using offline error to predict log-transformed online error for NNs that do not crash, and the corresponding Pearson's R-squared is shown in the top left-hand corner.}
 \label{fig:offlinevonline_noozone}
\end{figure}

\begin{figure}[!htbp]
 \centering
 \includegraphics[width=\textwidth]{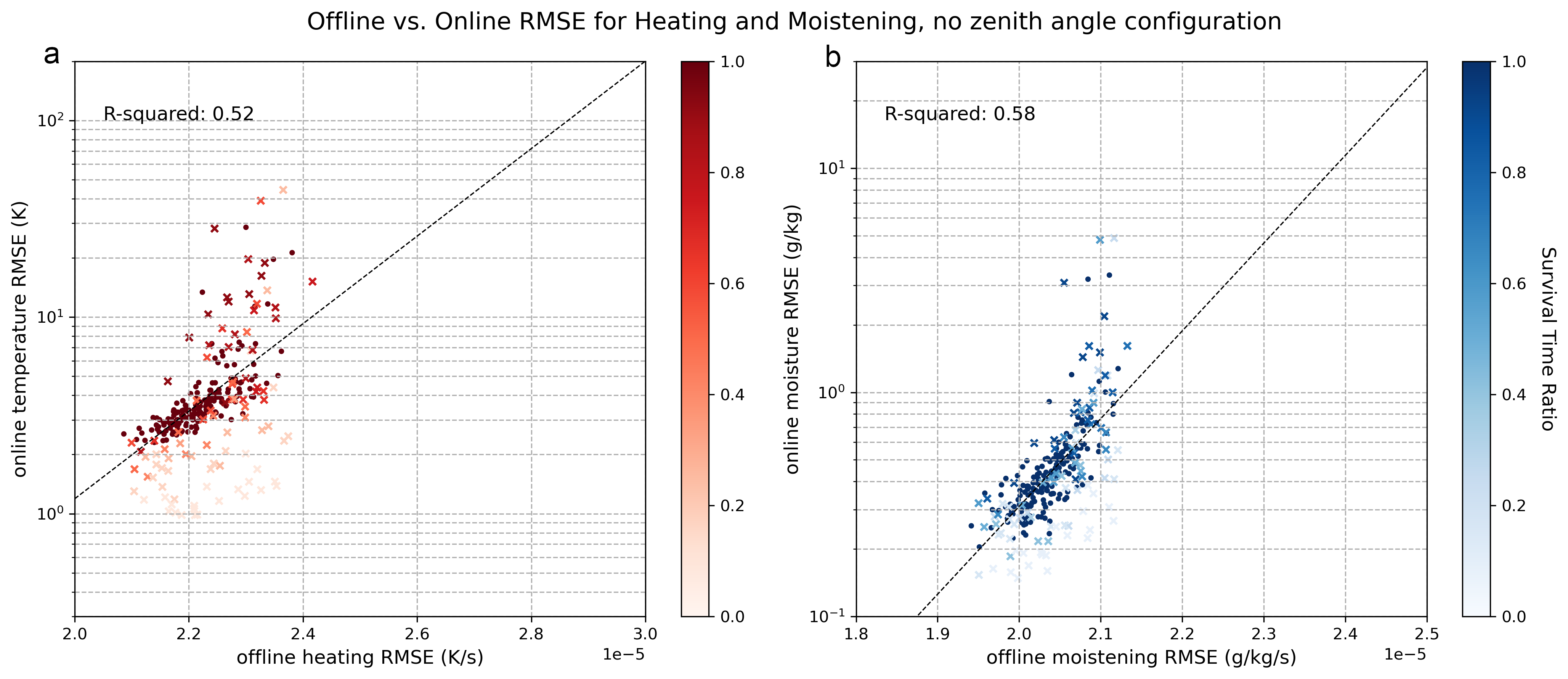}

\caption{This figure shows offline error vs. online error across all 330 models in the no zenith angle configuration. Models that crash mid-simulation are denoted with an `x'. A linear regression is fit using offline error to predict log-transformed online error for NNs that do not crash, and the corresponding Pearson's R-squared is shown in the top left-hand corner.}
 \label{fig:offlinevonline_nocoszrs}
\end{figure}

\begin{figure}[!htbp]
 \centering
 \includegraphics[width=\textwidth]{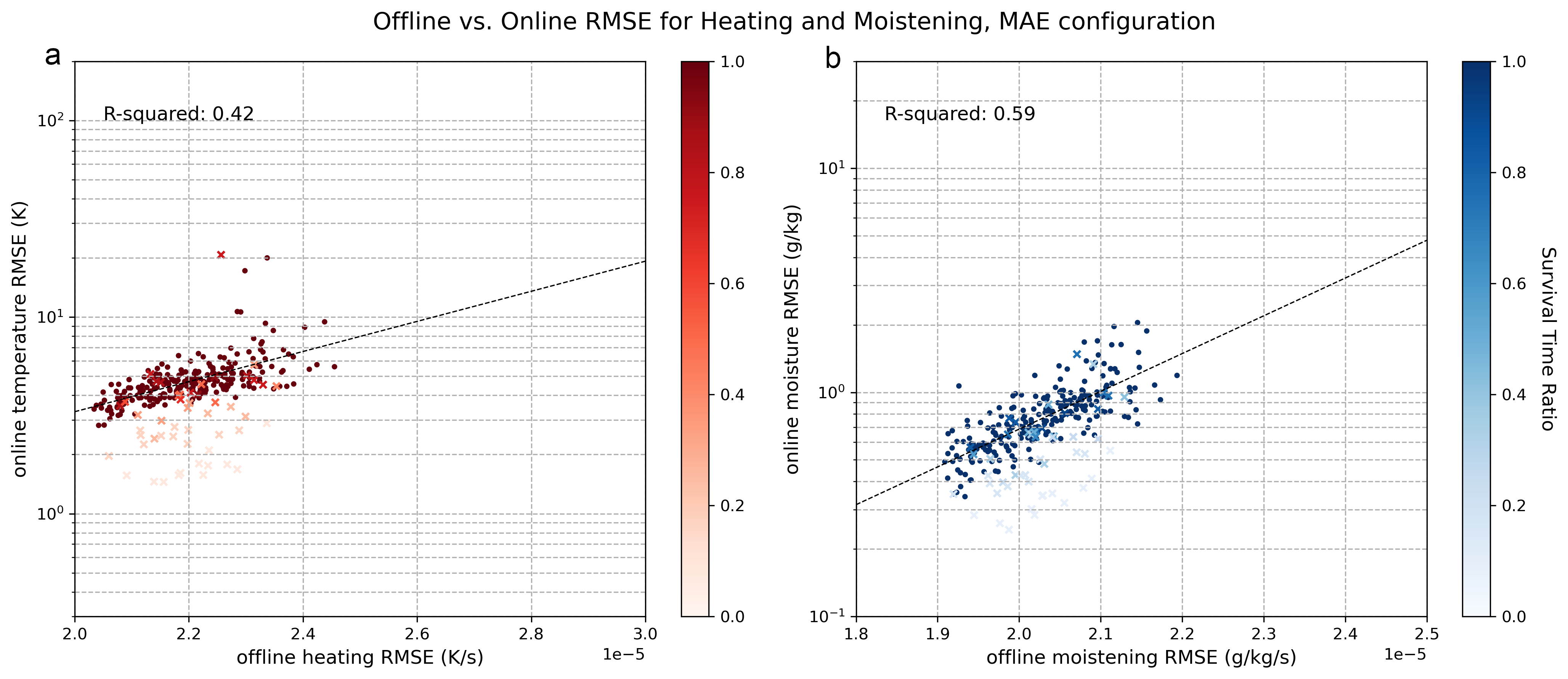}

\caption{This figure shows offline error vs. online error across all 330 models in the MAE configuration. Models that crash mid-simulation are denoted with an `x'. A linear regression is fit using offline error to predict log-transformed online error for NNs that do not crash, and the corresponding Pearson's R-squared is shown in the top left-hand corner.}
 \label{fig:offlinevonline_mae}
\end{figure}

\begin{figure}[!htbp]
 \centering
 \includegraphics[width=\textwidth]{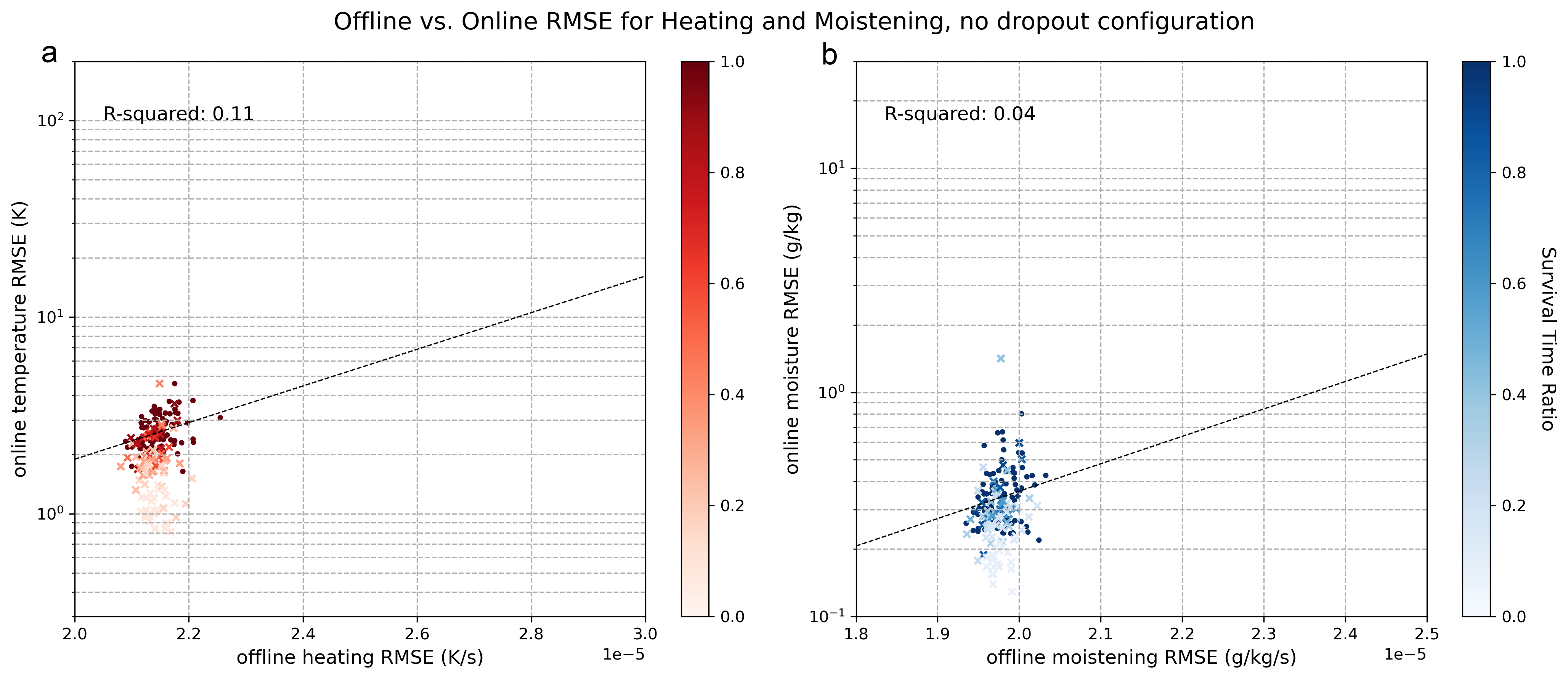}

\caption{This figure shows offline error vs. online error across all 330 models in the no dropout configuration. Models that crash mid-simulation are denoted with an `x'. A linear regression is fit using offline error to predict log-transformed online error for NNs that do not crash, and the corresponding Pearson's R-squared is shown in the top left-hand corner.}
 \label{fig:offlinevonline_nodropout}
\end{figure}

\begin{figure}[!htbp]
 \centering
 \includegraphics[width=\textwidth]{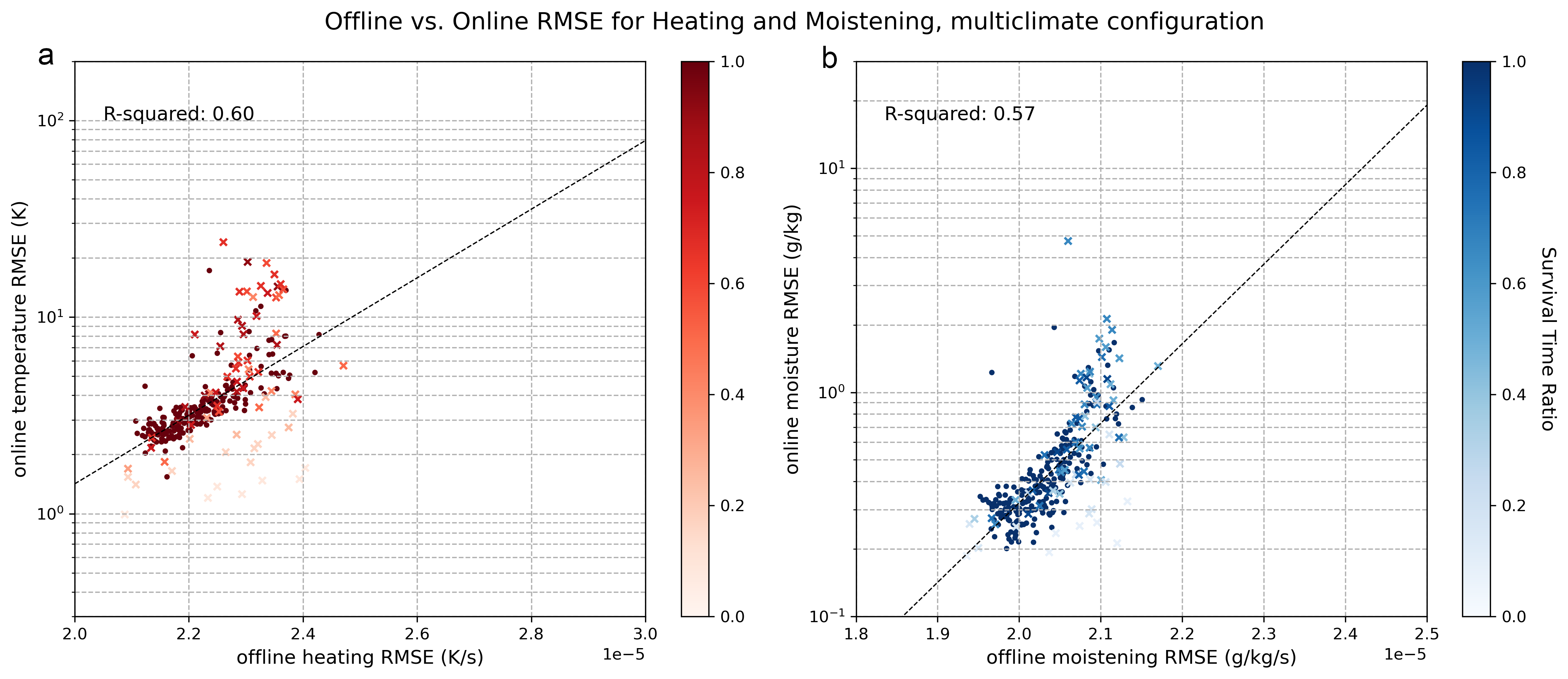}

\caption{This figure shows offline error vs. online error across all 330 models in the multiclimate configuration. Models that crash mid-simulation are denoted with an `x'. A linear regression is fit using offline error to predict log-transformed online error for NNs that do not crash, and the corresponding Pearson's R-squared is shown in the top left-hand corner.}
 \label{fig:offlinevonline_multiclimate}
\end{figure}

\begin{figure}[!htbp]
 \centering
 \includegraphics[width=\textwidth]{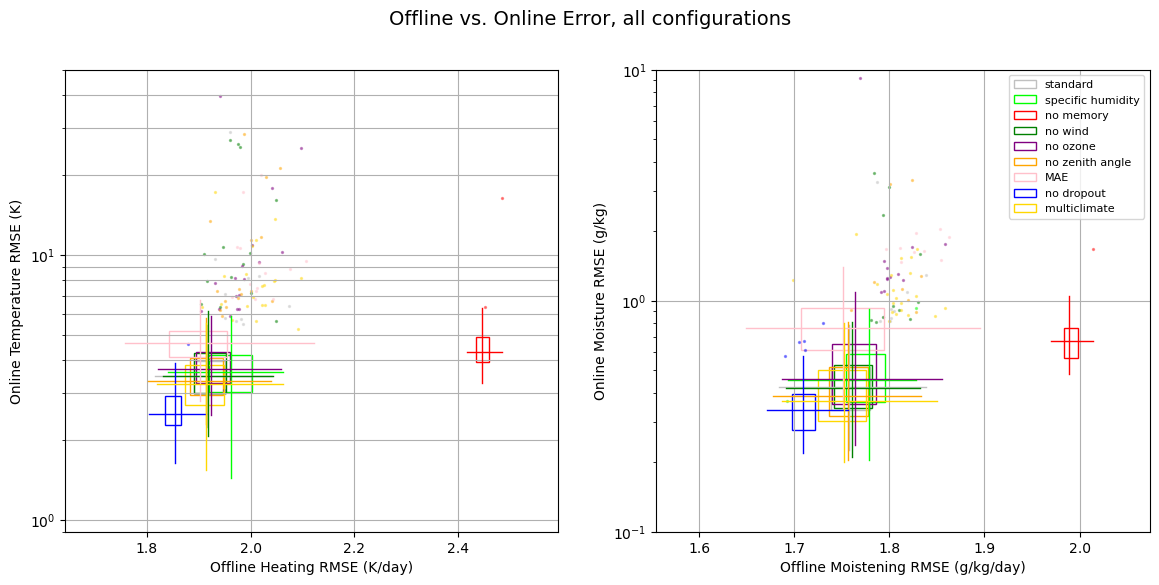}

\caption{This figure shows offline error vs. online error boxplots for all configurations. The Spearman correlation between offline heating error and online temperature error across all configurations is .549, and the corresponding spearman correlation for offline moistening error and online temperature is .531. These are both statistically significant.}
 \label{fig:offlinevonline_all}
\end{figure}

\begin{figure}[!htbp]
 \centering
 \includegraphics[width=\textwidth]{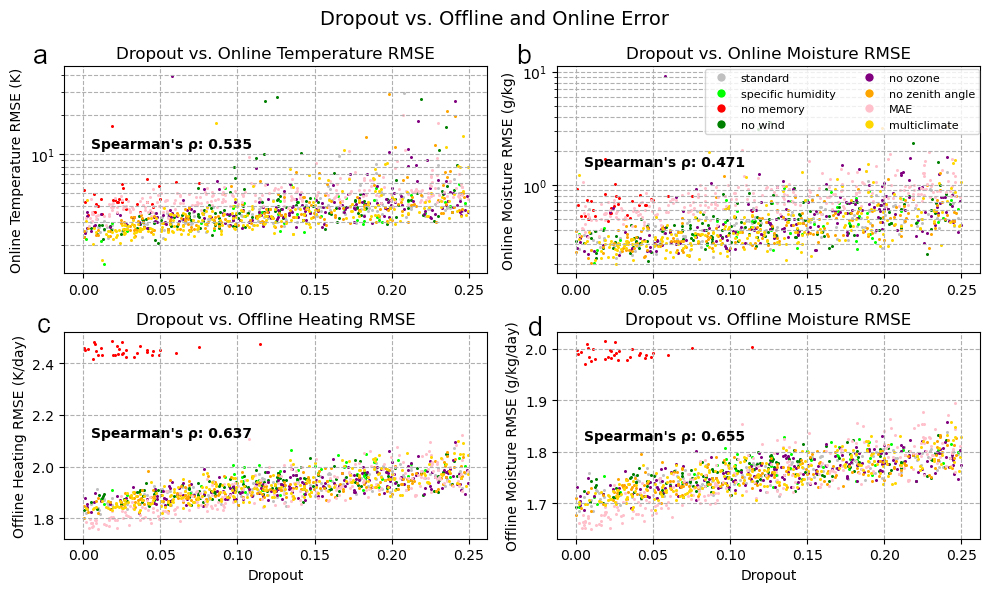}

\caption{This shows the effect of dropout on both offline heating (moistening) and online temperature (moisture) error after excluding the ``no dropout" configuration and marginalizing out choice of configuration. Bolded values are statistically significant. Offline and online error values for hybrid runs that crashed are excluded.}
 \label{fig:dropoutvserror_all}
\end{figure}

\begin{figure}[!htbp]
 \centering
 \includegraphics[width=\textwidth]{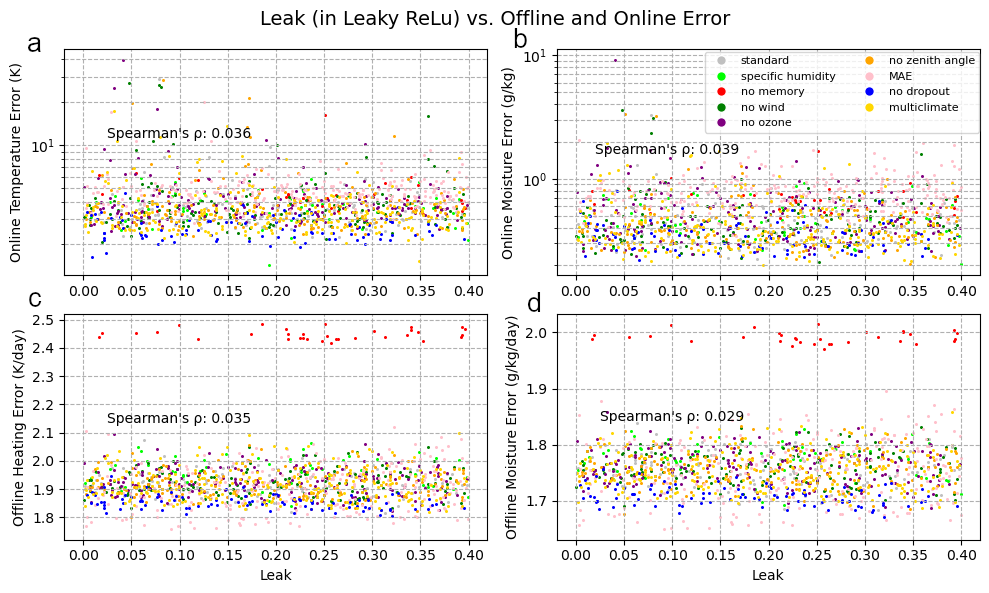}

\caption{This shows the effect of leak in Leaky ReLu on both offline heating (moistening) and online temperature (moisture) error after marginalizing out choice of configuration. Bolded values are statistically significant. Offline and online error values for hybrid runs that crashed are excluded.}
 \label{fig:leakvserror_all}
\end{figure}

\begin{figure}[!htbp]
 \centering
 \includegraphics[width=\textwidth]{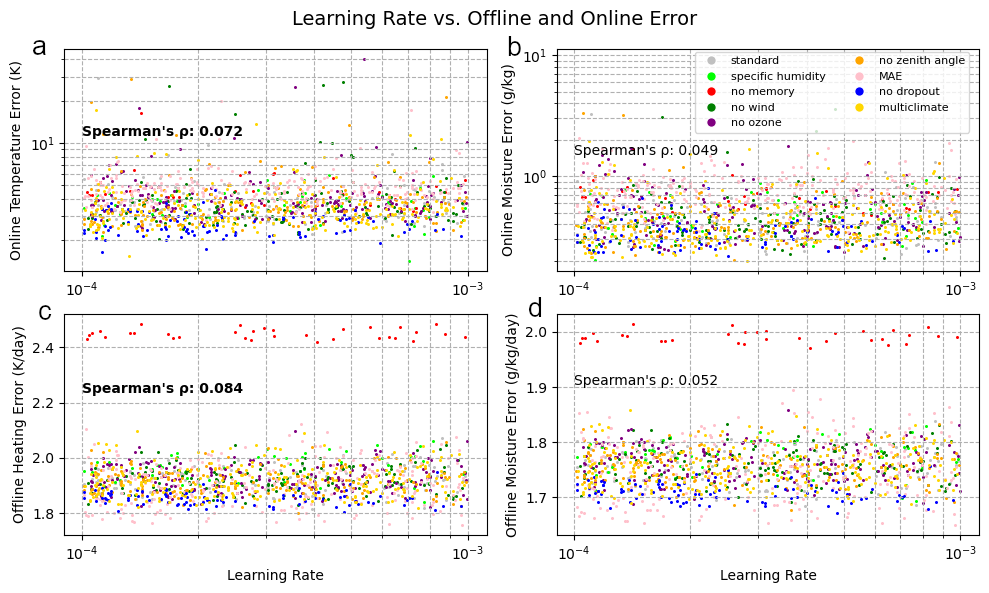}

\caption{This shows the effect of learning rate on both offline heating (moistening) and online temperature (moisture) error after marginalizing out choice of configuration. Bolded values are statistically significant. Offline and online error values for hybrid runs that crashed are excluded.}
 \label{fig:lrvserror_all}
\end{figure}

\begin{figure}[!htbp]
 \centering
 \includegraphics[width=\textwidth]{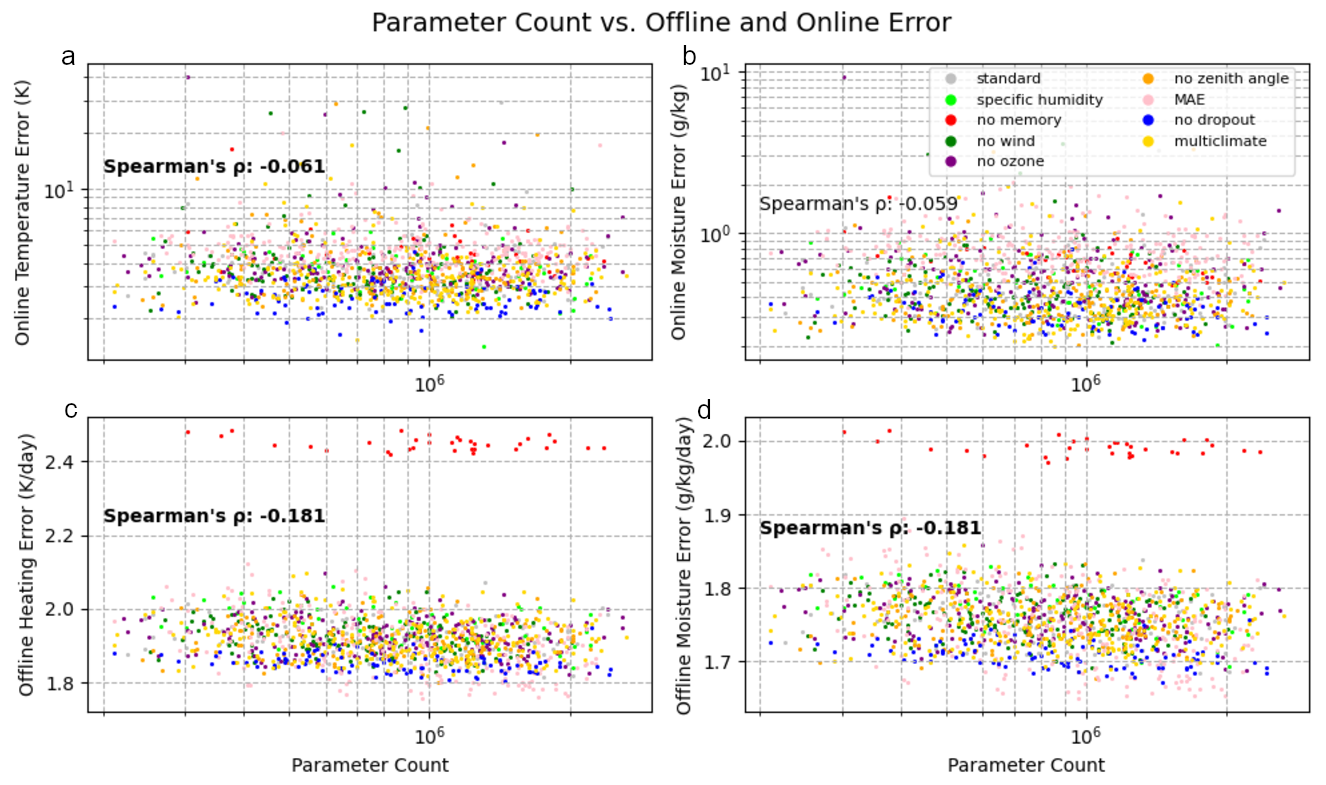}

\caption{This shows the effect of parameter count on both offline heating (moistening) and online temperature (moisture) error after marginalizing out choice of configuration. Bolded values are statistically significant. Offline and online error values for hybrid runs that crashed are excluded.}
 \label{fig:pcvserror_all}
\end{figure}

\begin{figure}[!htbp]
 \centering
 \includegraphics[width=\textwidth]{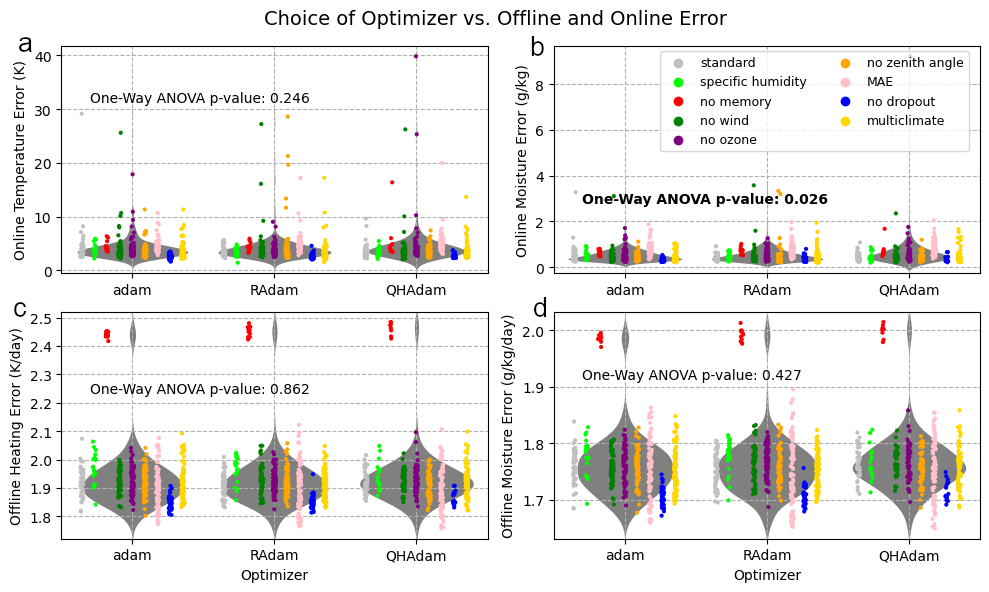}

\caption{This shows offline heating (moistening) and online temperature (moisture) error after conditioning on choice of optimizer and marginalizing out choice of configuration. We test for differences in offline and online error using one-way ANOVA tests. Although the one-way ANOVA test for online moisture error is significant, the subsequent Tukey's Honestly Significant Difference post hoc test did not uncover significant differences after correcting for the false discovery rate. Offline and online error values for hybrid runs that crashed are excluded.}
 \label{fig:optimizervserror_all}
\end{figure}

\begin{figure}[!htbp]
 \centering
 \includegraphics[width=\textwidth]{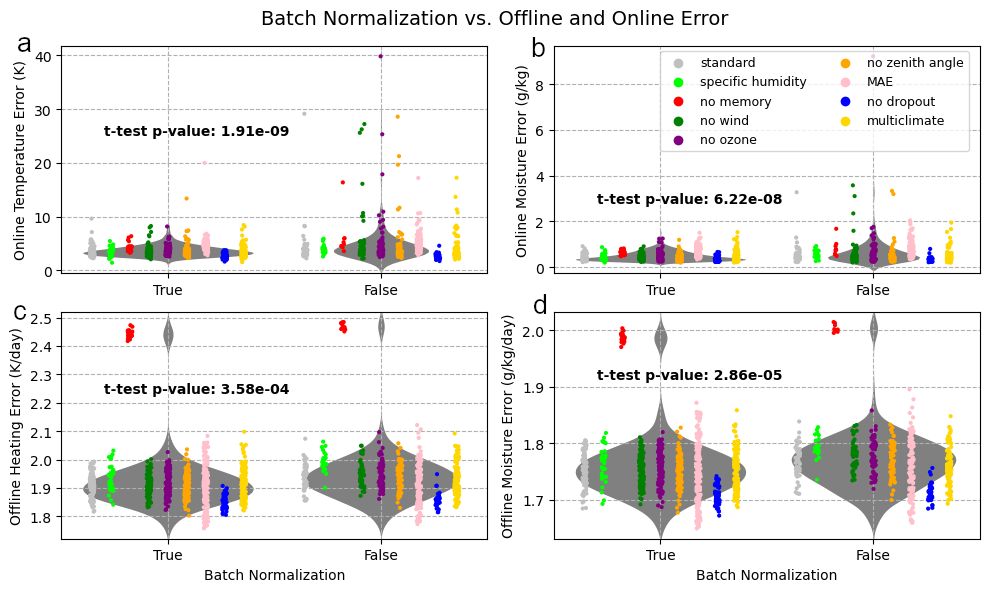}

\caption{This shows offline heating (moistening) and online temperature (moisture) error after conditioning on batch normalization and marginalizing out choice of configuration. We check for differences in mean offline and online error with and without batch normalization using two-tailed t-tests. Bolded values are statistically significant. Offline and online error values for hybrid runs that crashed are excluded.}
 \label{fig:bnvserror_all}
\end{figure}

\begin{figure}[!htbp]
 \centering
 \includegraphics[width=\textwidth]{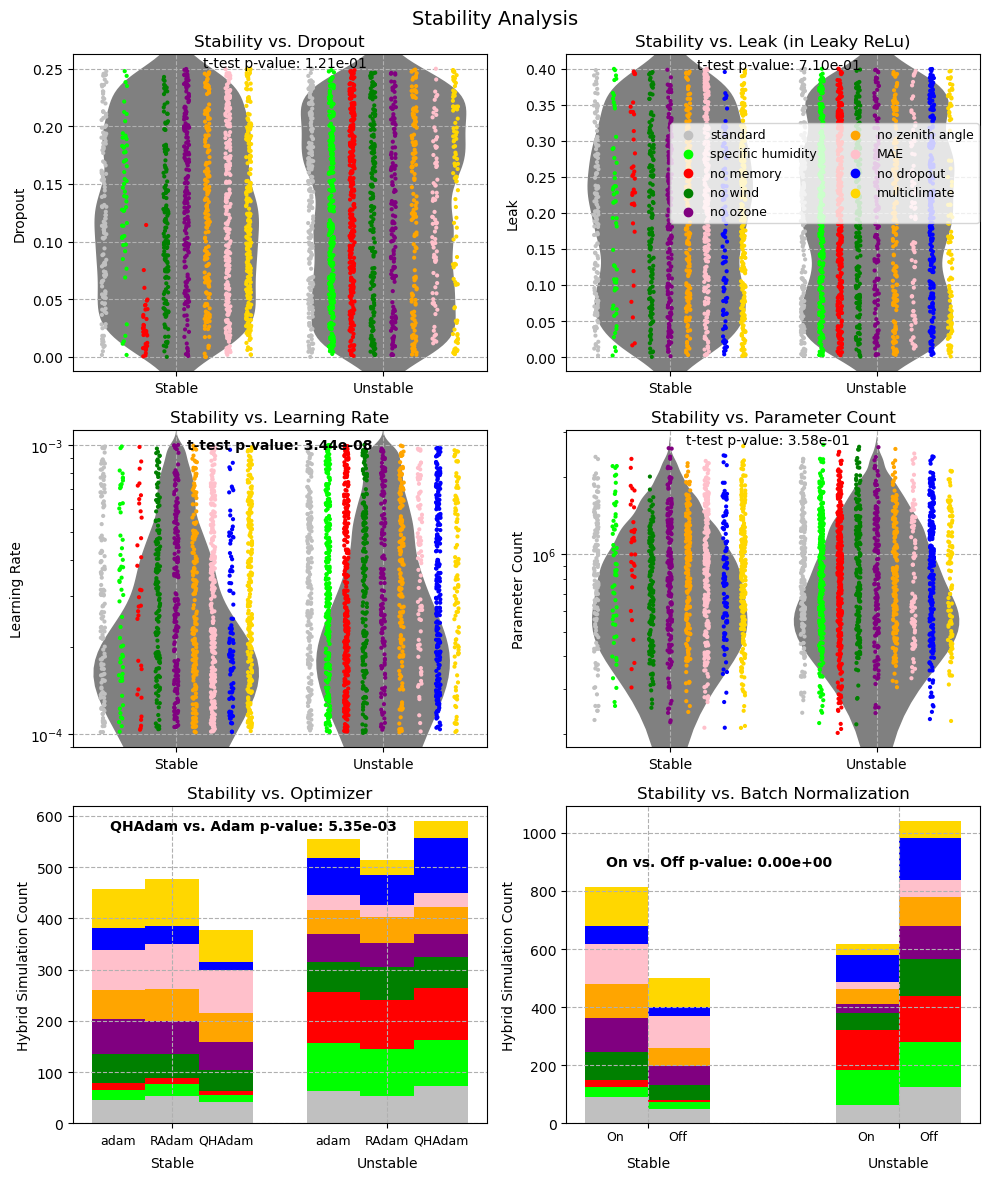}

\caption{This figure shows the effect of dropout, leak (in Leaky ReLu), learning rate, parameter count, and choice of optimizer on stability. We check for differences in mean dropout, leak, learning rate, and parameter count in stable and unstable NNs using two-tailed t-tests. For choice of optimizer and batch normalization, we make use of pairwise two-sample two-tailed proportion tests. RAdam is statistically equivalent to Adam, but QHAdam results in more unstable NNs than either. Bolded values are statistically significant.}
 \label{fig:stability_analysis}
\end{figure}

\begin{figure}[!htbp]
 \centering
 \includegraphics[width=\textwidth]{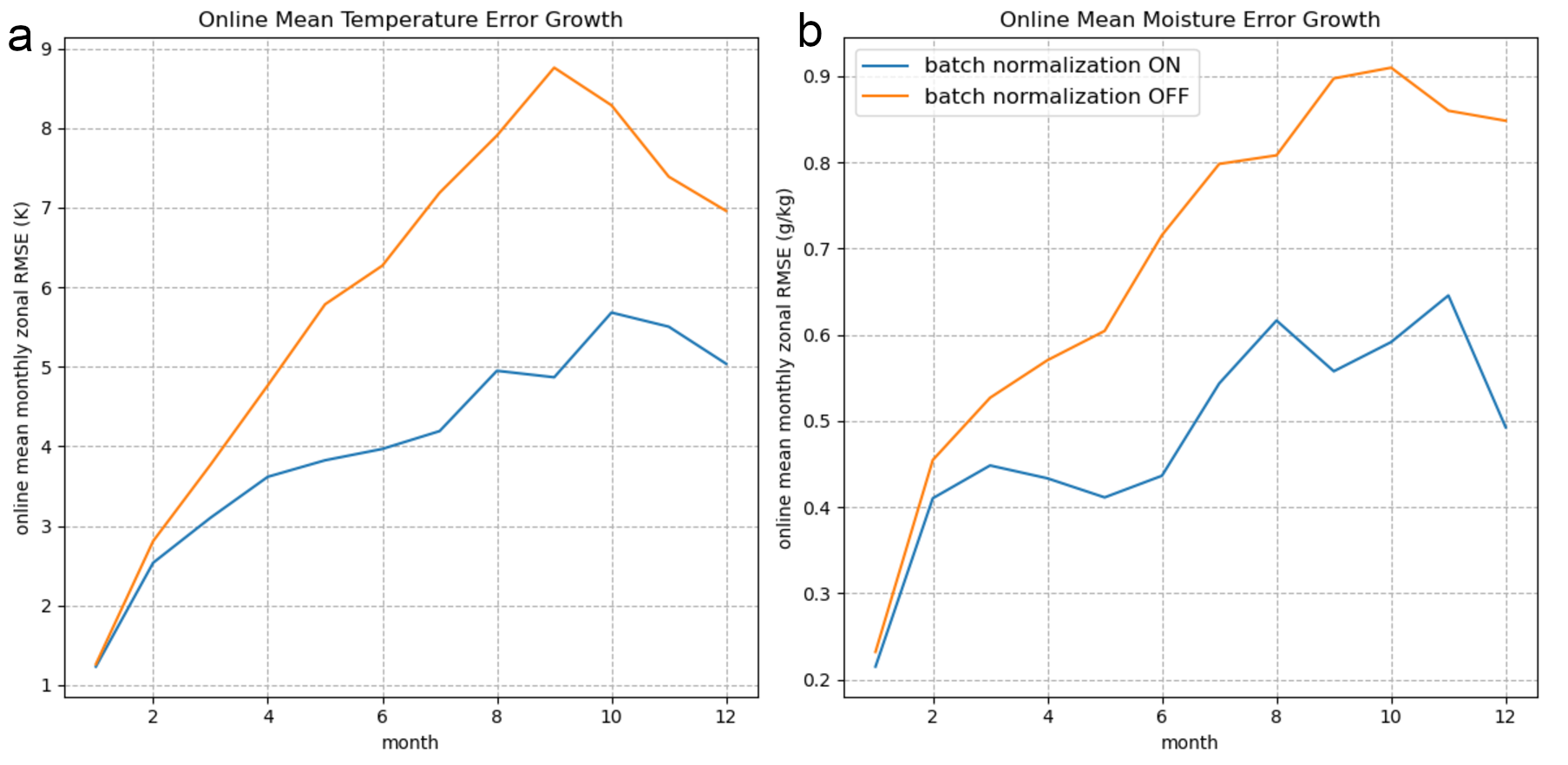}

\caption{This figure shows mean online error for both temperature and moisture over time for each hybrid simulations with and without batch normalization. Ensemble-mean online error is only recorded for simulations that do not crash, and they are calculated across the entire duration of the simulations.}
 \label{fig:bn_mean_error_growth}
\end{figure}

\begin{figure}[!htbp]
 \centering
 \includegraphics[width=\textwidth]{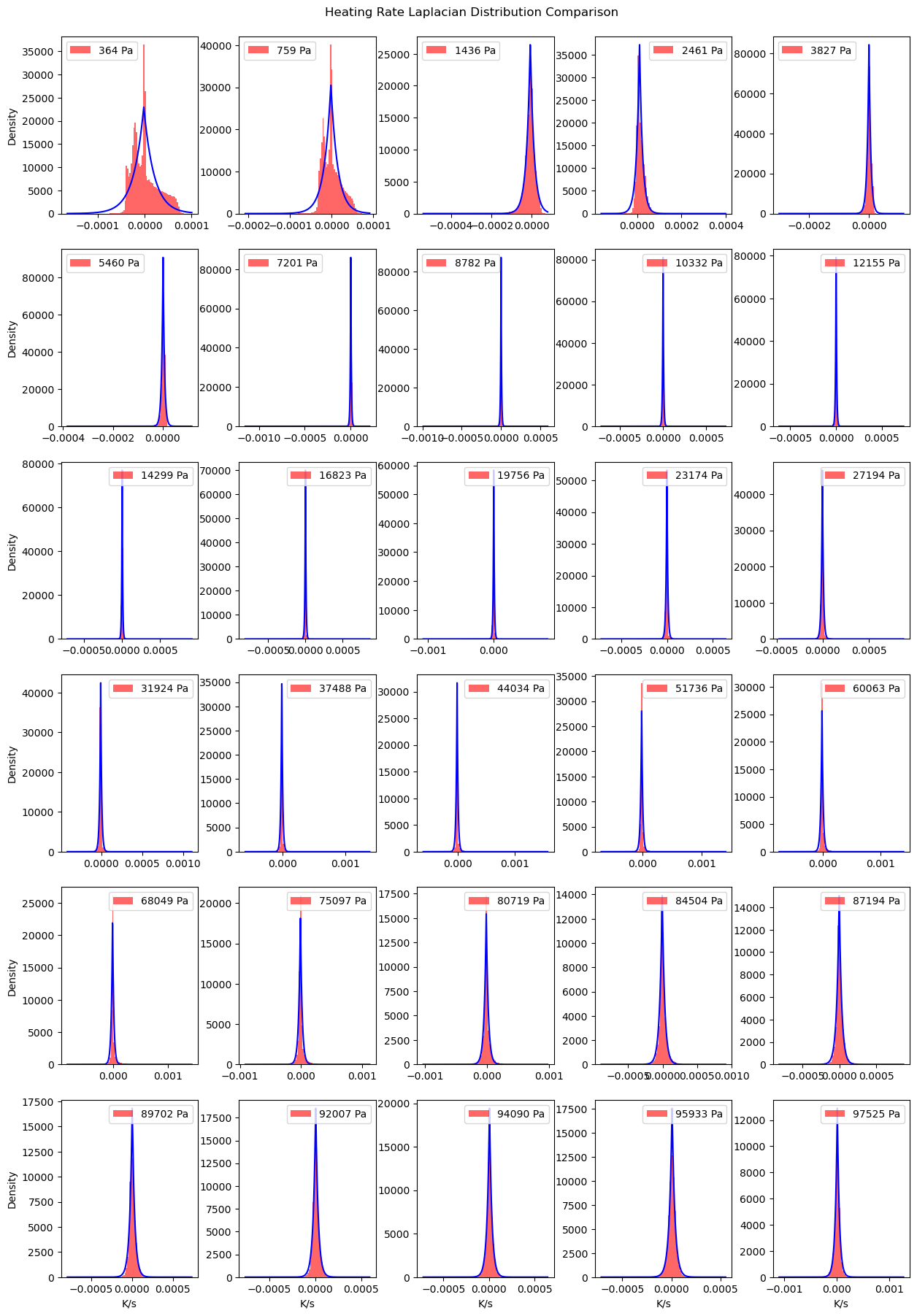}

\caption{This figure shows Laplacian distributions being fit to histograms of the heating tendencies at each vertical level.}
 \label{fig:heating_laplacian}
\end{figure}

\begin{figure}[!htbp]
 \centering
 \includegraphics[width=\textwidth]{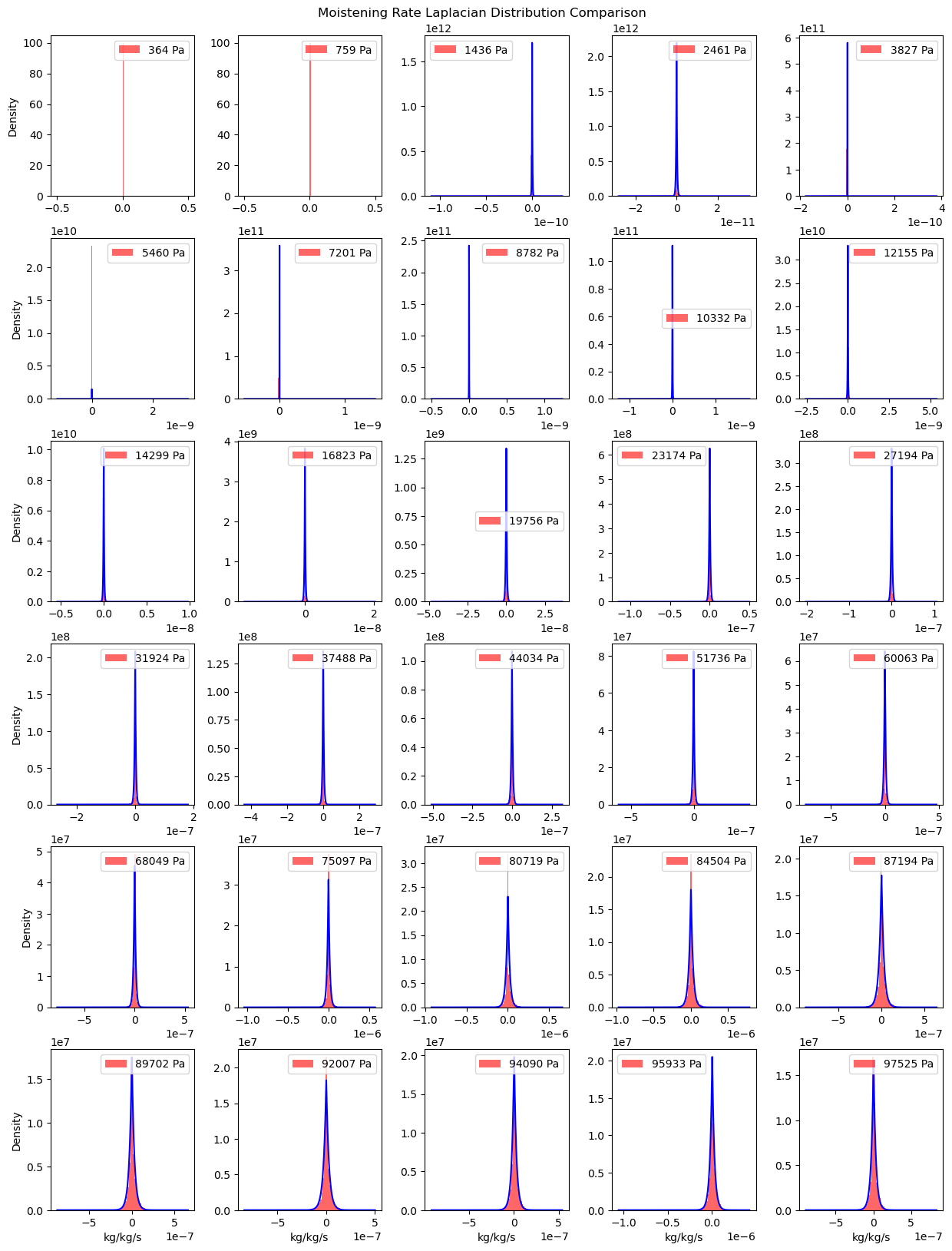}

\caption{This figure shows Laplacian distributions being fit to histograms of the moistening tendencies at each vertical level. The top five levels are included in this figure for completeness but are not used for the NN emulators.}
 \label{fig:moistening_laplacian}
\end{figure}

\end{document}